\documentclass[10pt,journal,compsoc]{IEEEtran}
\usepackage{cite}
\usepackage{amsmath,amssymb,amsfonts}
\usepackage{amsthm}
\usepackage{algorithmic}
\usepackage{graphicx}
\usepackage{textcomp}
\usepackage{xcolor}
\usepackage{url}
\usepackage{multirow}
\usepackage{hyperref}
\usepackage{caption}
\usepackage{subcaption}

\newtheorem{remark}{Remark}

\newcommand{\high}[1]{{\color{black}{#1}}}

\def\BibTeX{{\rm B\kern-.05em{\sc i\kern-.025em b}\kern-.08em
    T\kern-.1667em\lower.7ex\hbox{E}\kern-.125emX}}
\begin{document}

\title{GazePair: Efficient Pairing of Augmented Reality Devices Using Gaze Tracking}

\author{Matthew~Corbett,
        Jiacheng~Shang,
        and~Bo~Ji
\thanks{
The work of Jiacheng Shang was supported by NSF grant CNS 2153397.

Matthew Corbett (matthewc84@vt.edu) and Bo Ji (boji@vt.edu) are with the Department
of Computer Science, Virginia Tech, Blacksburg,
VA, 24061.
Jiacheng Shang (shangj@montclair.edu) is with the Department
of Computer Science, Montclair State University, Montclair, NJ 07043.
Bo Ji is the corresponding author.}
}

\maketitle

\begin{abstract}
As Augmented Reality (AR) devices become more prevalent and commercially viable, the need for quick, efficient, and secure schemes for pairing these devices has become more pressing. Current methods to securely exchange holograms require users to send this information through large data centers, creating security and privacy concerns. Existing techniques to pair these devices on a local network and share information fall short in terms of usability and scalability. These techniques either require hardware not available on AR devices, intricate physical gestures, removal of the device from the head, do not scale to multiple pairing partners, or rely on methods with low entropy to create encryption keys. To that end, we propose a novel pairing system, called GazePair, that improves on all existing local pairing techniques by creating an efficient, effective, and intuitive pairing protocol. GazePair uses eye gaze tracking and a spoken key sequence cue (KSC) to generate identical, independently generated symmetric encryption keys with 64 bits of entropy. GazePair also achieves improvements in pairing success rates and times over current methods. Additionally, we show that GazePair can extend to multiple users. Finally, we assert that GazePair can be used on any Mixed Reality (MR) device equipped with eye gaze tracking.
\end{abstract}

\section{Introduction}
\label{sec:intro}

According to recent market research, the global Augmented Reality (AR) market in 2021 was estimated at \$14.7 Billion, with an expected value of \$88.4 Billion by 2026~\cite{ARMarket}. As companies and users begin to explore the possibilities of the Metaverse and the future of human interaction~\cite{VRMarket, Metaverse}, AR is rapidly expanding past its current hardware limitations to even more ubiquitous use. These uses, such as in the windshields of automobiles~\cite{ARWindshield}, use by the military for soldier augmentation~\cite{ArmyHL2}, performing remote surgery and training healthcare workers~\cite{MRHealthcare}, educating children in the classroom~\cite{MREducation}, and visualizing logistics bottlenecks~\cite{MRUseCases}, portend a future where AR devices are used as a part of our normal, every-day lives. With the advent of the Metaverse, even the governments of large cities are turning to AR to ensure that their citizens have access to essential services through this emerging and potentially pervasive technology~\cite{SeoulMetaverse}.

As these AR devices grow in utility, use, and impact on daily lives, schemes to pair two or more of such devices will become even more important. The pairing of AR devices and sharing of experiences is at the core of the value of AR devices, allowing users to not only experience a synthetic augmentation of the physical world but to share these objects, normally known as holograms, with others. On the other hand, AR devices present unique challenges and opportunities for pairing. AR devices, especially head-mounted displays (HMDs), allow the user to interact with her physical environment while the headset places synthetic objects such as holograms into the user's perception of the physical world. HMDs also allow for integrated eye tracking, a feature deemed one of the most transformative to the technology~\cite{FutureEyeGaze}. These features allow for unique ways to generate the entropy required for secure pairing of AR devices, noted as 60+ bits of entropy in~\cite{MinimumEntropy}. This is in contrast to other mobile devices, such as smartphones, that have limited ways for users to interact for device pairing.

\subsection{The Importance of Local Pairing of AR Devices}
The importance of efficient pairing of AR devices is made evident in previous works. As originally described in the earliest work on AR-specific local pairing, Looks Good To Me (LGTM)~\cite{10.1145/2995289.2995295}, local sharing is a problem that deserves to be studied separately. Local sharing allows users to communicate without using large-scale data backbones, for-profit cloud services, or cellular connections.  It also gives users the freedom to decide to keep their data local and within a more closed sphere of control. Currently, for two users to pair their AR devices and securely share information such as holograms, the best working solution involves exchanging an alphanumeric authentication string or using centralized systems such as Microsoft Azure~\cite{ASA}. If focusing on local, bootstrapping methods of pairing, the alphanumeric string method is the only known, implemented method. These strings are error-prone and time-consuming to use~\cite{OOBUsabilitySecurity}. To alleviate this problem, recent research has created systems to use AR's spatial awareness capability, combined with the AR user's ability to interact with the physical environment, to efficiently pair two AR devices. LGTM, HoloPair~\cite{10.1145/3134600.3134625}, and Tap-Pair~\cite{10.1145/3374664.3375740} are all current research works that focus on using AR-specific technologies to pair such devices.  Each of these works presents a novel way to use wireless localization or holograms to authenticate a shared secret and secure communication paths without using Public Key Infrastructure (PKI) to create keys. However, none of these works implement or test methods of pairing more than two devices, and none of them explore a new and powerful technology, \emph{eye gaze tracking}, for AR device pairing. Additionally, their proposed pairing techniques are specific to AR. AR-specific gestures or technologies require that each of these solutions be deployed to AR devices only, greatly limiting the deployability and scope of the solutions (e.g., not applicable to Virtual Reality (VR) devices).
In light of this, \emph{it remains highly challenging to achieve a high level of entropy required for AR device pairing, while simultaneously creating a scalable, usable, and widely deployable solution.} 

\subsection{GazePair} 

\emph{We propose to use eye gaze tracking to create the entropy required for secure pairing, and the usability and scalability desired by users.} We adopt eye gaze tracking in our design for the following reasons. \emph{First}, harnessing eye gaze simply requires the user to direct their eyes, or look, at a target. It requires little explanation. \emph{Second}, eye gaze tracking is a relatively new method of user interaction with digital systems, and has been identified by some of the largest companies producing Mixed Reality (MR) devices (including AR and VR) as one of the technologies driving the future of this industry~\cite{SlugThesis,FutureEyeGaze,ForbesEyeGaze}. \emph{Third}, an AR user's eye gaze is nearly invisible to an outside observer. Most AR devices have a partially opaque visor concealing the user's eyes. This prevents easy, direct observation of the target of the user's gaze.  

Using eye gaze to generate the symmetric encryption keys required to pair devices, however, introduces unique challenges. \emph{First}, eye gaze detection and logging involves inherent error~\cite{s21062234}, even if small. As a result, the discretization of this data can be challenging. Eye saccades, the natural movement of the eye from point to point, eye fatigue, and even user inattentiveness make this and other techniques difficult to implement and discretize. 
\emph{Second}, the transition of eye gaze data to a symmetric encryption key is non-trivial. The system must not only be robust but also scalable (i.e., able to simultaneously pair more than two devices). 
\emph{Third}, eye gaze and iris/retinal data can be uniquely identifying and are a potential privacy risk if leaked accidentally. Such a system must protect user identity and unique biometric data. 
\emph{Fourth}, this system must use the created entropy to generate symmetric keys without requiring key authentication systems that limit scalability. We judge that minimum entropy must be 60+ bits in keeping with~\cite{MinimumEntropy}.

To address these challenges, we propose a novel pairing solution, \emph{GazePair}, which uses \emph{eye gaze tracking} to produce the 60+ bits of entropy required to pair devices securely. GazePair is a proven advancement in terms of usability and scalability over existing solutions and is also deployable on the breadth of emerging MR devices that use eye gaze tracking. GazePair uses a set of numerically-labeled holograms, randomly generated in 3D space, and a spoken \emph{key sequence cue (KSC)}, to pair AR devices and mitigate the threat of Man in the Middle (MITM) attacks. GazePair also does not need to publicly exchange discretization or error correction parameters and does not require recording or sharing gaze data that can jeopardize user privacy.
To prototype GazePair, we use the Microsoft HoloLens~2 AR HMDs to prove the hypothesis that efficient and secure local pairing is possible using eye gaze tracking as the primary entropy source for symmetric keys.
Currently, the Microsoft HoloLens~2 is the most advanced AR HMD on the market~\cite{HoloLens2}. It incorporates spatial awareness, eye gaze tracking, a relatively large field of view, and is completely wireless. While our prototype implementation requires Microsoft's Mixed Reality Toolkit (MRTK), we expect that the GazePair design is applicable to any MR device that supports eye gaze tracking. 

We summarize the contributions of this paper as follows: 
\begin{list}{\labelitemi}{\leftmargin=1.5em \itemindent=-0.0em \itemsep=.2em}
  \item To the best of our knowledge, this is the first work to show that eye gaze data generated in AR devices can be used to produce the 60+ bits of entropy required for efficient and secure pairing. 
  \item We design a novel system, GazePair, to locally pair AR devices using eye gaze tracking and a spoken KSC. GazePair improves on existing techniques in terms of usability, scalability, and security. Additionally, the design of GazePair is applicable to any MR device with eye gaze tracking.
  \item We implement a prototype system of GazePair using Microsoft HoloLens~2 AR HMDs. The prototype achieves a \emph{98.3\% success rate and an average of 9.02 seconds to complete a one-to-one pairing over 120 tests}. With three users, a one-to-two pairing takes \emph{an average of 12.58 seconds and has a 96.6\% success rate over 120 tests.} No existing work is proven to locally pair more than two AR devices or could claim to be deployable across the breadth of MR devices incorporating eye gaze tracking. 
\end{list}

We organize the rest of this paper as follows. We first review research work related to AR device pairing in Section~\ref{sec:works}.  Then, we address the design and prototype implementation of GazePair in Sections~\ref{sec:design} and \ref{sec:implementation}, respectively. In Section~\ref{sec:evaluation}, we present the experimental results of our user testing-based evaluation. Finally, we discuss the limitations of GazePair and future work in Section~\ref{sec:Discussion} and make concluding remarks in Section~\ref{sec:conclusion}.

\section{Related Work}
\label{sec:works}
We review the expanse of current work related to the topic of efficient AR device authentication and pairing and divide the body of work into five subtopics. These topics are: secure pairing of mobile devices (Section~\ref{sec:SecurePairingMobile}), pairing of AR devices (Section~\ref{sec:SecurePairingAR}), eye gaze in security applications (Section~\ref{sec:AuthEyeGaze}), and a final comparison of all applicable AR pairing solutions (Section~\ref{sec:ComparisonExistingResearch}).

\subsection{Secure Pairing of Mobile Devices}
\label{sec:SecurePairingMobile}

Numerous studies have been completed on how to intuitively and efficiently \emph{pair} mobile devices. Works such as~\cite{Kumar2009AliceMB} explain that the primary challenge to pairing these devices is authentication, or the ability to verify who the pairing partner is and what the user expects it to be. One method proposed and used widely is out-of-band (OOB) communication to verify the intended pairing partner. Such a method allows users to independently verify that the exchanged cryptographic material (e.g., the keys) has not been intercepted or manipulated by a malicious third party through MITM attacks. Various OOB channels, such as human physiology~\cite{10.1145/2994551.2994556, 10.1145/2422966.2422975}, ambient signals and energy~\cite{10.1145/1999995.2000016, 10.1145/3372297.3417288}, simultaneous tapping to attenuate the received signal strength~\cite{10.1145/3287079}, and device acceleration~\cite{7363362}, all attempt to solve the MITM problem. While novel, none of these protocols incorporate the unique and powerful capabilities of current AR devices or are otherwise inapplicable to AR uses. These protocols either focus on pairing multiple devices worn or used by a single user or simply are inefficient in terms of user interaction and success rate. To further expand on this point, imagine trying to ``shake'' two AR headsets together to synchronize accelerometer data as in~\cite{7363362}. This would require users to remove their AR devices, if worn, something not desirable in constant, immersive AR environments. Other techniques, such as Apple's AirDrop~\cite{AirDrop} or Wi-Fi direct~\cite{WiFiDirect}, are difficult to scale and are designed for one-to-one connections. None of these techniques are particularly well-suited to the pairing of AR devices, but they are useful to understand as we seek to create an intuitive AR-friendly pairing protocol.

\subsection{Pairing of AR Devices}
\label{sec:SecurePairingAR}
The topic of efficient pairing of AR devices is most relevant to GazePair. While the body of work on this specific topic is still relatively small, three known techniques exist. First, Looks Good to Me (LGTM)~\cite{10.1145/2995289.2995295} is the first research done into using the inherent functionality of AR devices, such as the ability to physically see other users in an augmented space while wearing a device, to assist in secure pairing.  LGTM uses facial recognition and wireless localization to authenticate shared keys and to pair AR devices. The wireless localization hardware has never been deployed to any actual AR devices and LGTM cannot be effectively evaluated against any proposed pairing solution, including GazePair. LGTM also relies on facial recognition, tested using sample faces from the Yalefaces dataset without obstruction. Modern AR devices, such as the Microsoft HoloLens series of devices~\cite{HoloLens2}, partially obscure the human face, greatly frustrating this assumption. Even so, LGTM integrates AR-friendly technologies such as facial recognition and certainly breaks ground on efficient methods to pair AR devices.

HoloPair~\cite{10.1145/3134600.3134625} proposes a system that generates a secret locally on one device, creates a public key, transmits the key to a pairing partner, and authenticates this partner's key through a system of OOB physical interaction. HoloPair requires two users to trace the outline of a hologram created from the shared keys for key validation, and wave as a way to prevent MITM attacks during the transmission of the keys. Additionally, as identified in follow-up works by the authors~\cite{10.1145/3374664.3375740}, HoloPair allows users to simply accept the hologram verification, regardless of accuracy, potentially compromising the security of the protocol. This would potentially be done as a way to ``speed up'' the pairing by users. Additionally, HoloPair makes no claims that its protocol is feasible for use in larger than one-to-one simultaneous pairing scenarios. Finally, given the tracing requirement, HoloPair is an AR-only system, as users must see each other. However, HoloPair's reported pairing times, on average, are between 8-9 seconds. This is certainly an efficiency goal that pairing solutions should aspire to meet or exceed.

The most current work, Tap-Pair~\cite{10.1145/3374664.3375740}, improves on HoloPair by removing some of the elaborate requirements for user interaction. Tap-Pair also requires two co-located users, both wearing AR devices, to ``tap'' on the same physical location using a head direction to select a location. The original HoloLens used as the basis for Tap-Pair did not have true eye gaze tracking, but only the ability to sense the direction of the user's head. This ``tap'' requires users to direct their head at the desired point. This action gives a potential eavesdropping attacker insight into the shared secret. For this reason, the authors do not assume that an attacker can be located with legitimate users. Additionally, Tap-Pair does not allow for windows to be used as the physical location, limiting use cases. Unlike HoloPair, Tap-Pair claims that it is scalable to multiple users. However, the authors did not attempt to either implement or test such a solution, leaving its feasibility purely theoretical. Tap-Pair is also AR-specific, and users must see the physical environment. Still, Tap-Pair remains the most feasible and practical method for locally pairing two AR devices, specifically two (1st generation) HoloLens devices. 

\begin{table*}
\begin{tabular}{ p{1.6cm}||p{3.5cm}|p{1.75cm}|p{1.75cm}|p{1.8cm}|p{2.75cm}|p{2cm}}
 \hline
 \textbf{Method} & \textbf{Interaction 
 Technique} & \textbf{Success Rate} & \textbf{Pairing Time} & \textbf{Scalability} & \textbf{Security} & \textbf{Device} \\
 \hline
 \hline
 LGTM~\cite{10.1145/2995289.2995295} & No out-of-band communication & 58.4\% & Not reported & Unproven & Relies on wireless localization, entropy not reported & AR only \\
 \hline
 HoloPair~\cite{10.1145/3134600.3134625} & Tracing \& waving & 98\% & 10-11 seconds & Unproven  & Vulnerable to inattentiveness, 53 bits of entropy &  AR only   \\
 \hline
Tap-Pair~\cite{10.1145/3374664.3375740} & Head direction and tapping & 90\% & Not reported & Unproven & 9-11 bits of entropy & AR only \\
 \hline
  \textbf{GazePair (this work)} & Spoken key sequence cue \& eye gaze & 98.3\% & 9.02 seconds & 2+ user pairing proven & 64 bits of entropy  & MR with eye gaze tracking \\
 \hline
\end{tabular}
\caption{ A comparison of GazePair with state-of-the-art AR device pairing solutions.}
\label{tab:researchComparison}
\end{table*}

\subsection{Eye Gaze in Security Applications}
\label{sec:AuthEyeGaze}
While we know of no work using eye gaze to \emph{pair} two or more MR devices, there are numerous works that address using eye gaze to \emph{authenticate} a single user. Systems such as~\cite{RubikAuth} ensure authentication by asking a VR user to manipulate a 3D cube, similar to a Rubik's cube, and enter a password based on number and color using the hand-held controllers and the user's gaze. Other examples such as~\cite{GazeTouchPass} use a combination of eye gaze and user touchpad input for authentication. Regardless of the system, implementations such as these commonly cite two things: that eye gaze can be more secure, but also suffers from increased input times and error rates over more standard input methods. As an example, \cite{LookIntoMyEyes} shows that eye gaze input methods can be nearly 25-fold more time intensive than manually inputting a standard PIN on a keypad with failure rates of nearly 24\%. These cautionary tales guide us toward a solution that maximizes the security qualities of eye gaze, but in a system that does not use intricate eye gestures and can tolerate minor miscalibration of an eye gaze sensor.

\subsection{Comparison with State-of-the-Art Research}
\label{sec:ComparisonExistingResearch}
In each of the discussed AR pairing solutions, users are required to be wearing an AR device (and only an AR device). Additionally, no solution has been proven to be usable to pair more than two devices. All three existing systems require users to see each other or interact with the physical environment, limiting the system to AR applications. Also, other methods in the field of authentication exist, such as~\cite{InvestigatingTheThirdDimension}. Like Tap-Pair, these systems leverage the device user's understanding of the physical and digital world in order to create or validate some secret. However, this technique is shown to have low entropy (about 12 bits) and high failure rates similar to Tap-Pair. Like two of the three existing solutions, GazePair also requires an OOB communication channel for pairing. However, GazePair only requires that the pairing initiator speak a key sequence cue of an arbitrary length, something that requires much less motion than the tracing/waving or walking/pointing required from HoloPair and Tap-Pair respectively. We improve on these solutions by designing, implementing, and testing GazePair, a novel pairing system that is user-friendly and scalable. GazePair extends to more than two users and uses eye gaze for entropy generation, a technique that simply requires users to direct their eyes at a holographic target. Eye gaze requires very little movement from the users, is partially obscured from bystanders due to the opaque visor on most AR devices, and is shown during evaluations to be easy to learn and understand. Table~\ref{tab:researchComparison} presents a comparison of our proposed GazePair system with these existing AR pairing solutions. 

\section{GazePair Design}
\label{sec:design}

In this section, we propose a novel pairing system, \emph{GazePair}, which leverages \emph{eye gaze tracking}, a new and powerful AR technology.
GazePair improves on the shortcomings of existing pairing strategies discussed in Section~\ref{sec:SecurePairingAR}.
In GazePair, we assume that users are wearing AR HMDs with gaze-tracking sensors. These AR HMDs have access to a local network, but may or may not have access to the larger internet through this connection.  We also assume that the users must not be required to remove these AR HMDs to conduct any of the required tasks for pairing their devices. Additionally, we assume that the users are co-located, in such proximity that a spoken word can be easily heard by all legitimate clients. In the following, we first introduce the threat model in Section~\ref{sec:threat}, present the design goals in Section~\ref{sec:goals}, discuss the key challenges of using eye gaze tracking for pairing in Section~\ref{sec:challenges}, and elaborate on the design of GazePair in Section~\ref{sec:narrative}.

\begin{figure*}
     \centering
     \begin{subfigure}{0.45\textwidth}
         \centering
         \includegraphics[width=0.7\textwidth]{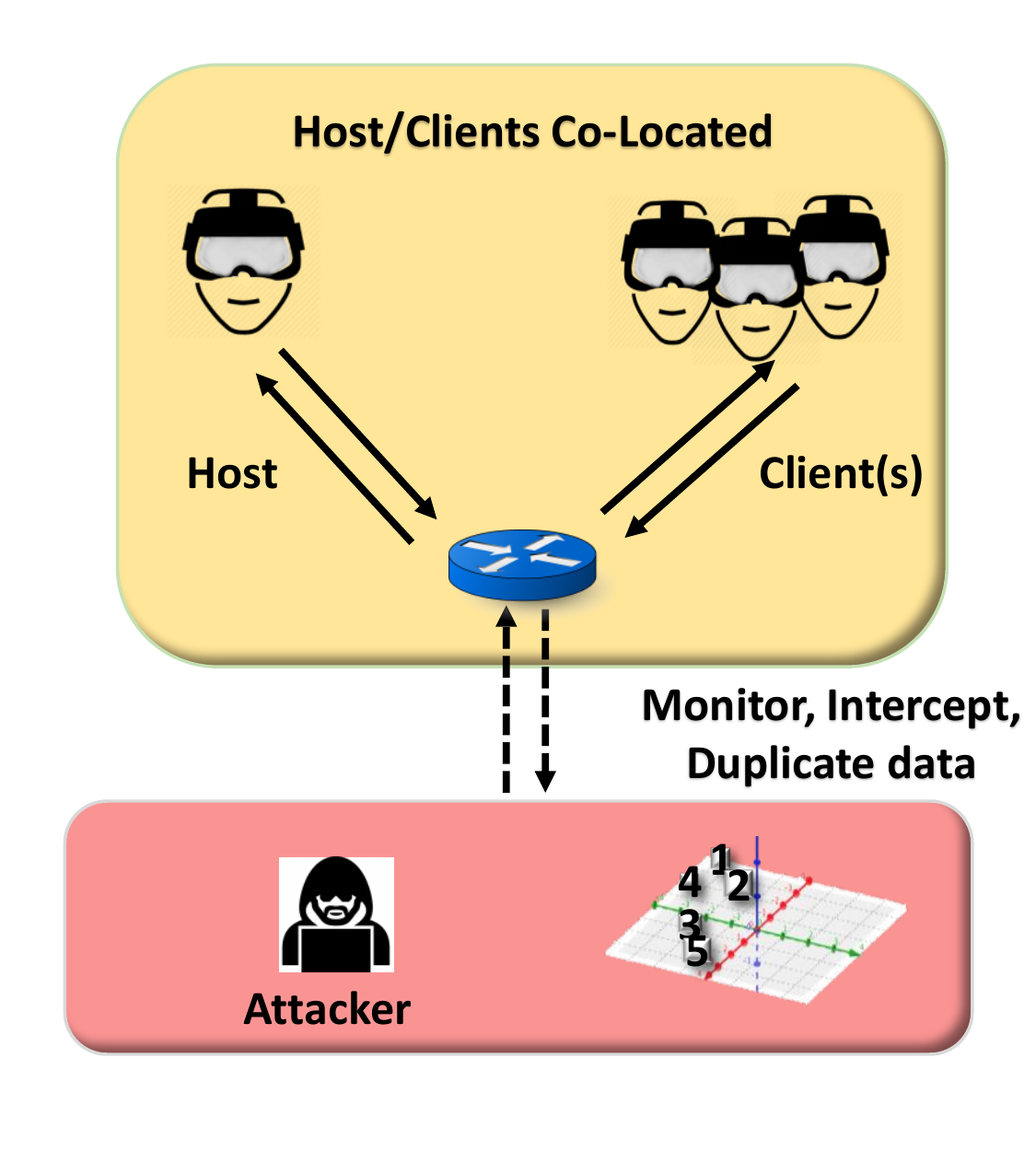}
         \caption{Attacker can observe shared holograms}
         \label{fig:AttackStrategiesa}
     \end{subfigure}
     \begin{subfigure}{0.45\textwidth}
         \centering
         \includegraphics[width=0.7\textwidth]{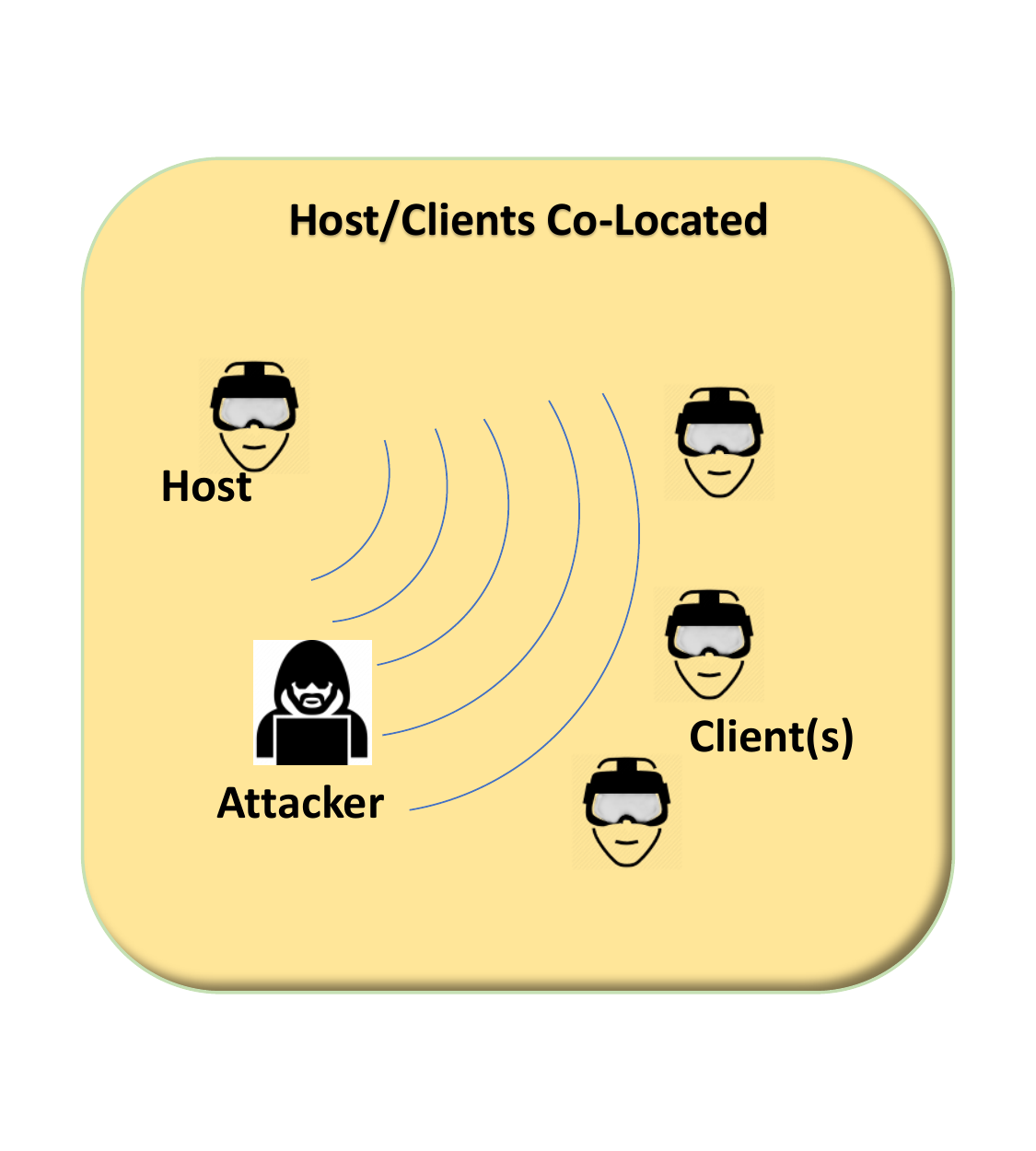}
         \caption{Attacker can eavesdrop on the vocal communication}
         \label{fig:AttackStrategiesb}
     \end{subfigure}
     \hfill
        \caption{ If the attacker has access to the local network, as illustrated in (a), the attacker has the ability to intercept traffic at will and can understand the location of any holograms presented to each user. The attacker cannot, however, overhear or eavesdrop on any activity in the physical proximity of the users. If the attacker has access to the physical location of the pairing, as illustrated in (b), the attacker has access to the vocal communication between the pairing partners, but not the network traffic being passed amongst the pairing partners. Here, the attacker may have knowledge of what is said or done among the pairing partners, but does not have a way to use the passing network traffic to exploit this information.}
        \label{fig:AttackStrategies}
\end{figure*}

\subsection{Threat Model}\label{sec:threat}
For GazePair, we assume a threat model that is similar to those proposed in the most recent research on AR device pairing~\cite{10.1145/2995289.2995295, 10.1145/3134600.3134625, 10.1145/3374664.3375740}. We assume that this threat has access to the local network that the legitimate users are using to conduct the pairing, and may monitor, intercept, and duplicate packets arbitrarily. The threat may also be physically located with the users, something additional to current research on the topic. However, we assume that the attacker \emph{cannot} do both simultaneously, such as intercepting packets on the local network while also being within earshot of the legitimate users. The threat may also have access to an AR device (e.g., a HoloLens~2), but does not need it to intercept the initially shared holograms. We believe that this is realistic. One bit of logic behind this assumption is that it is either similar to those used in existing pairing techniques~\cite{10.1145/2995289.2995295} or more advanced. Other techniques consider only that the attacker can be physically present~\cite{10.1145/3134600.3134625} \emph{or} that they can have access to network traffic~\cite{10.1145/3374664.3375740}, but do not consider these scenarios in the development of the same technique.
Finally, we do not consider Denial of Service (DoS) attacks in this research. Such DoS attacks include limiting access to the local network, physically interrupting the attempted pairing, and destroying traffic for the sole purpose of disrupting pairing. Attack strategies are depicted in Fig.~\ref{fig:AttackStrategies}.

\subsection{Design Goals }\label{sec:goals}
In order of priority, we design GazePair with five principles in mind. 

\begin{list}{\labelitemi}{\leftmargin=1.5em \itemindent=-0.0em \itemsep=.2em}
    \item \textbf{Usability.} The GazePair user interface and required operations must be intuitive to novice AR device users. GazePair must not require the use of any advanced interaction techniques to navigate the user interface. The completion of a GazePair pairing must not take an excessive amount of the user's time.
    
    \item \textbf{Success rate.} GazePair must be effective, meaning that user pairing success rate must be 95\% or higher.
    
    \item \textbf{Scalability.} GazePair must allow more than a one-to-one pairing and must be able to complete these pairings without excessive additional time requirements. Since keys generated from this pairing are symmetric, GazePair must be able to efficiently generate multiple matching keys.
    
    \item \textbf{Security.} GazePair must produce an acceptable level of security through the level of entropy from the shared secret created by harnessing eye gaze. We believe this level to be more than 60 bits of entropy~\cite{MinimumEntropy}. GazePair must also be resistant to MITM attacks.
    
    \item \textbf{Device requirement.} GazePair should operate across the breadth of MR devices that incorporate eye gaze tracking, and allow hardware-independent, efficient pairing for each. GazePair must be deployable on any such MR device quickly and with no software package conflicts, by simply changing the intended deployment platform in Unity.
\end{list}

We propose that eye gaze tracking can be harnessed to achieve each of the design goals above while improving on the shortcomings of current works noted in Section~\ref{sec:SecurePairingAR}. However, there are unique challenges with using eye gaze for this purpose. They are outlined in Section~\ref{sec:challenges} and are addressed in our GazePair design.

\subsection{Key Challenges}
\label{sec:challenges}

As mentioned in Section~\ref{sec:intro}, using eye gaze tracking for both user input and entropy generation involves inherent error. The underlying mechanisms to harness this gaze data once collected are non-trivial. Additionally, the user's uniquely identifying eye and gaze-tracking data must be protected. In the following, we discuss these key challenges in detail.

\textbf{Challenge 1: Eye gaze tracking error.}  
Eye gaze tracking has inherent error. Both research works and Microsoft's eye tracking documentation note a roughly 1.5 to 3-degree variance in expected gaze location compared to the recorded gaze location~\cite{EyeTrackingHL2, s21062234}. This error only increases when the user moves and dictates the minimum size of eye gaze targets with the user stationary or in motion.

\textbf{Challenge 2: The requirement for OOB communication.}  
The ability of an attacker to intercept and exploit the location of a shared hologram led us to believe that OOB communication was necessary, similar to the authors of HoloPair~\cite{10.1145/3134600.3134625} and Tap-Pair~\cite{10.1145/3374664.3375740}. Additionally, direct communication with intended pairing partners also provides built-in authentication, similar to existing works as well.  

\textbf{Challenge 3: Eye gaze for symmetric key generation.}
While gaze data has been heavily studied for the purposes of authentication in MR devices~\cite{10.1145/3489849.3489880, 10.1145/3204493.3208333, 10.1145/3379157.3391421, 10.1145/2976749.2978311, 10.1145/1280680.1280683, 10.1145/1117309.1117319}, it has yet to be studied for the pairing of AR devices in order to exchange information securely. To this point, gaze data has been identified as a rapidly increasing research area for securing Human-Computer Interaction (HCI)~\cite{10.1145/3313831.3376840}, but to our knowledge, mechanisms to create symmetric keys from eye gaze data do not yet exist. A system must be created that can take a user's eye gaze, discretize as required, and use this secret information to create a symmetric encryption key. This system must be robust enough to ensure a reasonable amount of security without creating a burden to the user.

\textbf{Challenge 4: Data privacy.} Eye gaze data can uniquely identify a user and has the potential to harm users if the collected data is misused. As such, techniques have been developed to remove the uniquely identifying traits from gaze data, if and when it needs to be transmitted in insecure environments~\cite{9382914}. A system using eye gaze tracking must ensure that the uniquely identifying biometric data is not leaked or transmitted in such a way as to jeopardize the user's privacy.  

GazePair overcomes these challenges and creates a usable, scalable, and efficient pairing system using eye gaze. GazePair mitigates inherent eye tracking data by constructing holograms to meet the design specifications in~\cite{s21062234} and creates a discretization mechanism to compensate for gaze tracking error. Additionally, GazePair implements a method to transform the gaze data into a shared secret and creates a symmetric encryption key to be used to protect further user communication across a local network. A spoken sequence cue is used to direct the users to the correct numbered holographic keys and is used as the OOB channel.  The design of GazePair and the underlying pairing protocol are detailed in Section~\ref{sec:narrative}.

\begin{figure*}[!t]
\centering
\includegraphics[scale=0.7]{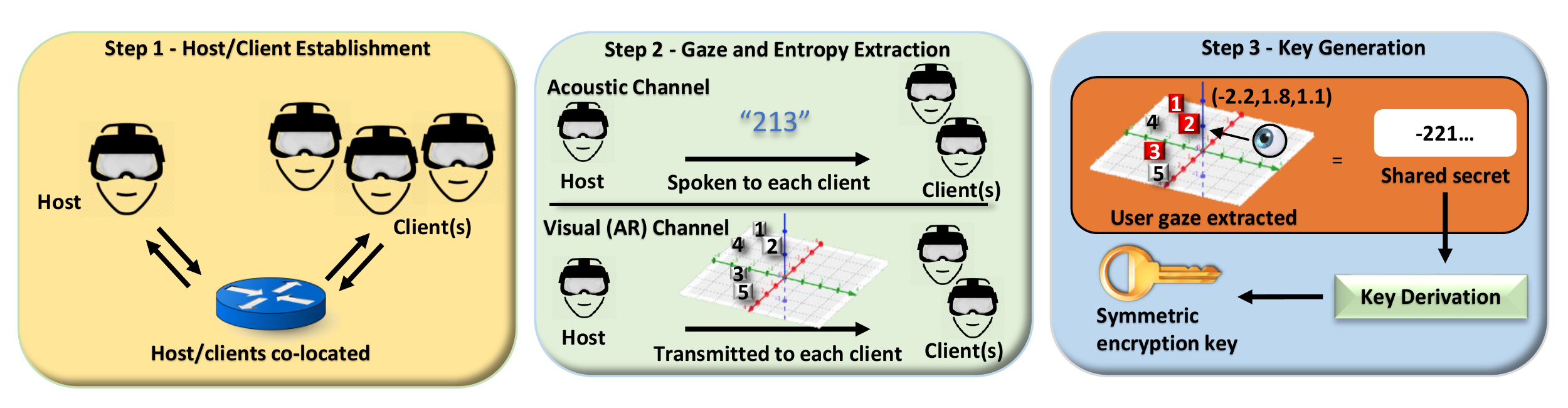}
\caption{An illustration of GazePair's three main steps to key generation and device pairing.} 
\label{fig:Overview}
\end{figure*}

\subsection{Pairing Process of GazePair}
\label{sec:narrative}
In order to create the symmetric encryption key required to complete the pairing of the AR devices, GazePair operates in three main steps as shown in Fig.~\ref{fig:Overview}. We explain these steps as follows:

   \textbf{Step~1: Host/Client Establishment.} GazePair creates a host/client relationship on a local network, e.g., through a User Datagram Protocol (UDP) connection. This is necessary to ensure that information (i.e., numerically-labeled holograms) are accurately shared between two or more users.
   
   \textbf{Step~2: Gaze and Entropy Extraction.} The host randomly places ten numerically-labeled holograms (denoting digits from 0 to 9) within a certain visibility threshold. The size of the keys is determined by the eye gaze tracking error in \emph{Challenge~1}, based on pilot testing. Specifically, this size is a compromise between large keys limiting generated entropy and small keys limiting the accuracy of eye gaze. The host then speaks\footnote{The KSC can also be written if users have significant obstacles to speaking and/or hearing the KSC.} a \emph{key sequence cue (KSC)} to each client. This process addresses the OOB communication challenge noted in \emph{Challenge~2}. Each client selects the required numbers in sequence, dictated by the KSC, one at a time using eye gaze. The shared secret becomes the concatenated, discretized values of the 3D location of the user's gaze when the user selects each numerically-labeled hologram. This process harnesses the user's eye gaze to obtain the entropy needed for creating symmetric encryption keys, while not requiring storage of any data that can compromise user privacy, which addresses \emph{Challenges~3 and 4}.
   
    \textbf{Step~3: Key Generation.} Key generation uses the shared secret to create a symmetric encryption key, e.g., an Advanced Encryption Standard (AES) key. This key is a function of the shared secret created by the gaze tracking data. Hence, the entropy of this shared secret must be as large as possible to create strong keys.

\begin{figure*}[!t]
\centering
\includegraphics[scale=.65]{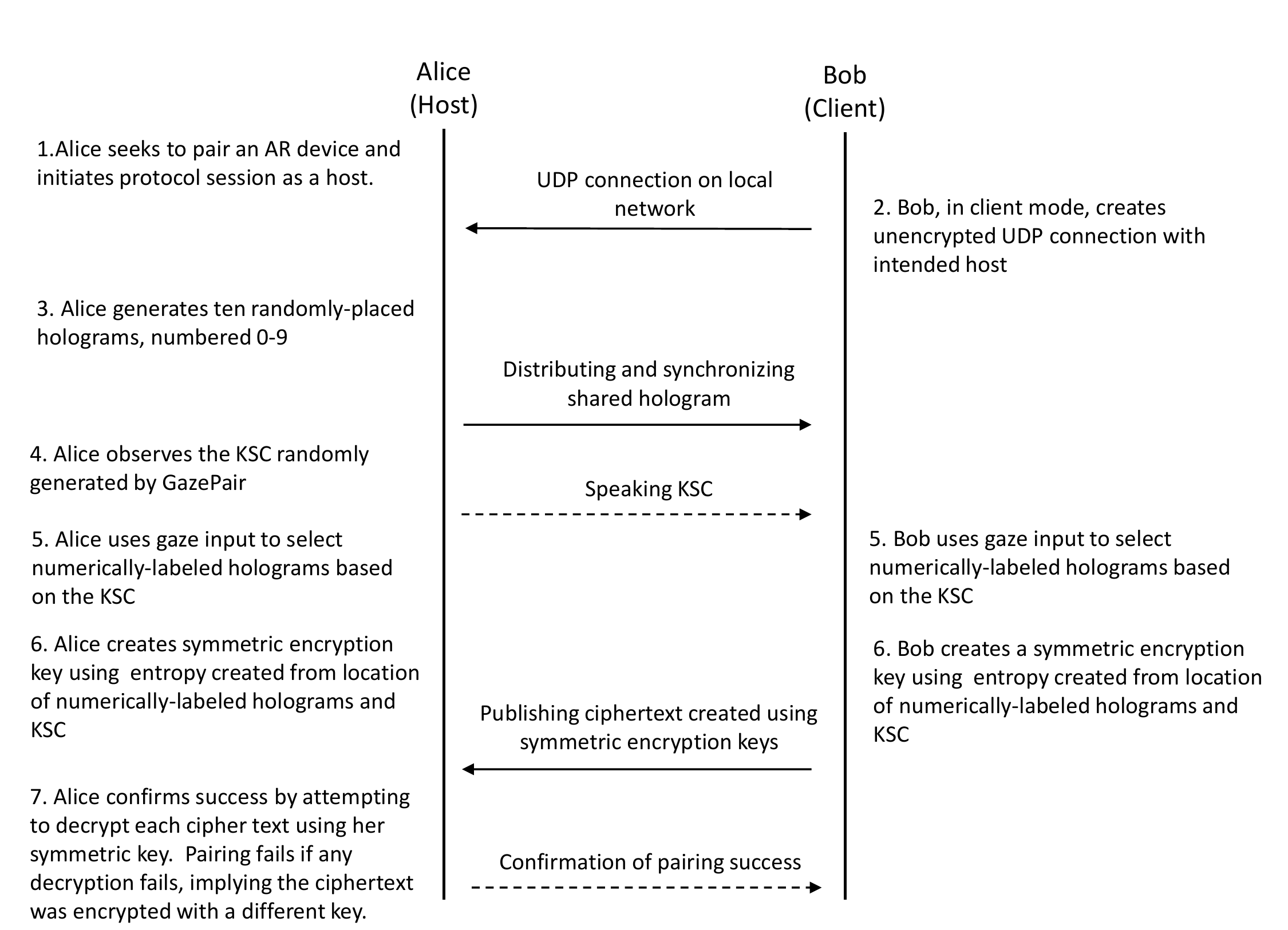}
\caption{A depiction of the pairing protocol, using the classic Alice and Bob analogy. While there is only a single client (Bob) in this illustration, we show that GazePair is scalable to multiple clients easily. 
} 

\label{fig:Protocol}

\end{figure*}

Next, we describe the pairing protocol.
As shown in Fig.~\ref{fig:Protocol}, a GazePair session begins with the host selecting their role in the initial GazePair connection scene (Step~1).  This allows a client or clients to initiate the protocol appropriately, and establish an unencrypted UDP connection to the host (Step~2). With this connection established, the host randomly generates ten numerically-labeled holograms, numbered 0-9 (Step~3). These numerically-labeled holograms are randomly placed in the augmented 3D space, with limits. These limits are designed to ensure that the holograms are within a given visibility threshold. These holograms are placed relative to the position of each pairing participant and are not anchored to any one physical location. The client then receives these locations and reproduces the set of numerically-labeled holograms on the client's HMD. Then, the host observes the randomly generated KSC (Step 4) and speaks the KSC to the client. Each participant uses gaze input to select numerically-labeled holograms in sequence based on the KSC (Step 5).

If the gaze ray impact point is inside the discretization range of any hologram, that hologram turns red to indicate to the user the key has been successfully entered. The discretized values of the gaze position are then concatenated to generate the shared secret string independently on each HMD. Consider the example in Fig.~\ref{fig:Overview}. Suppose the KSC is ``213''. For the first numerically-labeled hologram selected (i.e., digit ``2'', the first digit of the KSC), a gaze location of $(-2.2,1.8,1.1)$ is captured and might be corrected to the string ``-221'' based on discretization parameters, and this becomes the first three digits of the shared secret. This continues until the required KSC length is met, and the protocol waits for each participant to complete this process. After completing the entry, each HMD independently creates a symmetric encryption key (Step 6) and encrypts a message using this key. The pairing protocol succeeds if and only if the host can decrypt each client's ciphertext (Step 7). 

\begin{remark}
GazePair uses a KSC similar to a padlock combination or a keypad on a door. On the face of the fact, this sounds similar to a traditional passcode-based pairing instance and may seem insecure or trivial. However, the KSC is of no value to any attacker without detailed knowledge of the location of the numerically-labeled holograms in 3D space.
\end{remark}

\begin{remark}
It may seem that the use of gaze tracking to select the numerically-labeled holograms is extraneous.  It is certainly possible to modify GazePair to use other standard AR (or generally MR) gestures and inputs. For instance, using the ``hand ray'' gesture, creating a simple ray from a user's extended hand, can be used to select the numerically-labeled holograms in sequence based on the KSC. While it is possible to do this, we lose the benefit of the obscuration of the user's eyes. With the HoloLens~2, the user's eye direction is greatly obscured by the plastic visor upon which the holograms are projected. Without this, an attacker could more easily understand the location that the user is intending to select, aiding in deciding on the location and sequence of the numerically-labeled holograms being detected. This vulnerability has been well-documented in works such as~\cite{9417659}, where researchers use channel state information to conduct side-channel key logging attacks against VR users with hand held controllers. To prevent this, GazePair uses the "air tap" gesture which does not require the gesture to be oriented toward the intended target in any way, unlike the "hand ray" gesture~\cite{AirTap, HandRay}. We elaborate on the air tap gesture's utility in~Remark~\ref{rem:airtap}.
\end{remark}

\section{Prototype Implementation}
\label{sec:implementation}
Using the GazePair design described above, we then seek to create a usable prototype to prove our hypothesis. Using the Unity development environment, Microsoft's Mixed Reality Toolkit (MRTK), and multiple HoloLens 2 AR HMDs, we create a working prototype of GazePair that advances the state-of-the-art with proven usability and scalability. GazePair creates 64 bits of total entropy in the shared secret while not exposing the user's uniquely identifying eye data to potential misuse. Additionally, GazePair is deployable on any MR device capable of eye gaze tracking. 
In the following, we first briefly introduce the Microsoft's MRTK (Section~\ref{sec:MRTK}), and then describe the detailed implementation of GazePair prototype in three main steps: Host/Client Establishment (Section~\ref{sec:HostClientEstablishment}), Gaze and Entropy Extraction (Section~\ref{sec:GazeEntropyExtraction}), and Key Generation (Section~\ref{sec:KeyGeneration}).

\subsection{Microsoft's Mixed Reality Toolkit}
\label{sec:MRTK}

Created and maintained by Microsoft, the Mixed Reality Toolkit (MRTK), is an attempt to standardize the development of MR applications~\cite{MRTK}. MRTK is at the core of the HoloLens, and its successor, the HoloLens~2. Additionally, it is used to design applications for Meta's Oculus series of MR devices, as well as HTC's Vive, and other Windows Mixed Reality headsets. Most importantly to this research, it also contains the APIs used to access the HoloLens~2's powerful eye-tracking technology. As we begin to explore the possibility of using eye gaze to pair multiple MR devices, the deployability of this prototype on MR devices depends on the use of MRTK for eye tracking, but the design itself is usable on any MR device incorporating eye gaze tracking.  At this moment, only the HoloLens~2 incorporates eye tracking using MRTK, but as this expands, so does any pairing solution created with MRTK's eye-tracking APIs, including any and all MR devices using this specific capability.

\subsection{Host/Client Establishment}
\label{sec:HostClientEstablishment}

\subsubsection{Mid-level API (MLAPI)}
In order to distribute these numerically-labeled holograms, a networking suite must be used to create and synchronize these holograms. For GazePair, we choose a Unity-developed API called Mid-level API (MLAPI) to create this functionality. MLAPI is open source and simply seeks to abstract transport layer functionality to ease the integration of networking functionality into Unity-created applications. MLAPI allows the user to easily declare their intent to initiate a session as a host or a client, and function accordingly. While MLAPI communicates to clients without encryption, the information we share publicly (i.e., the location of numerically-labeled holograms) is not sufficient to breach the security of GazePair.  After confirmation of matching symmetric Advanced Encryption Standard (AES) 256-bit keys, users can use these keys to encrypt any communication desired.

\begin{figure*}[!t]
     \centering
     \begin{subfigure}{0.26\textwidth}
         \centering
         \includegraphics[width=\textwidth]{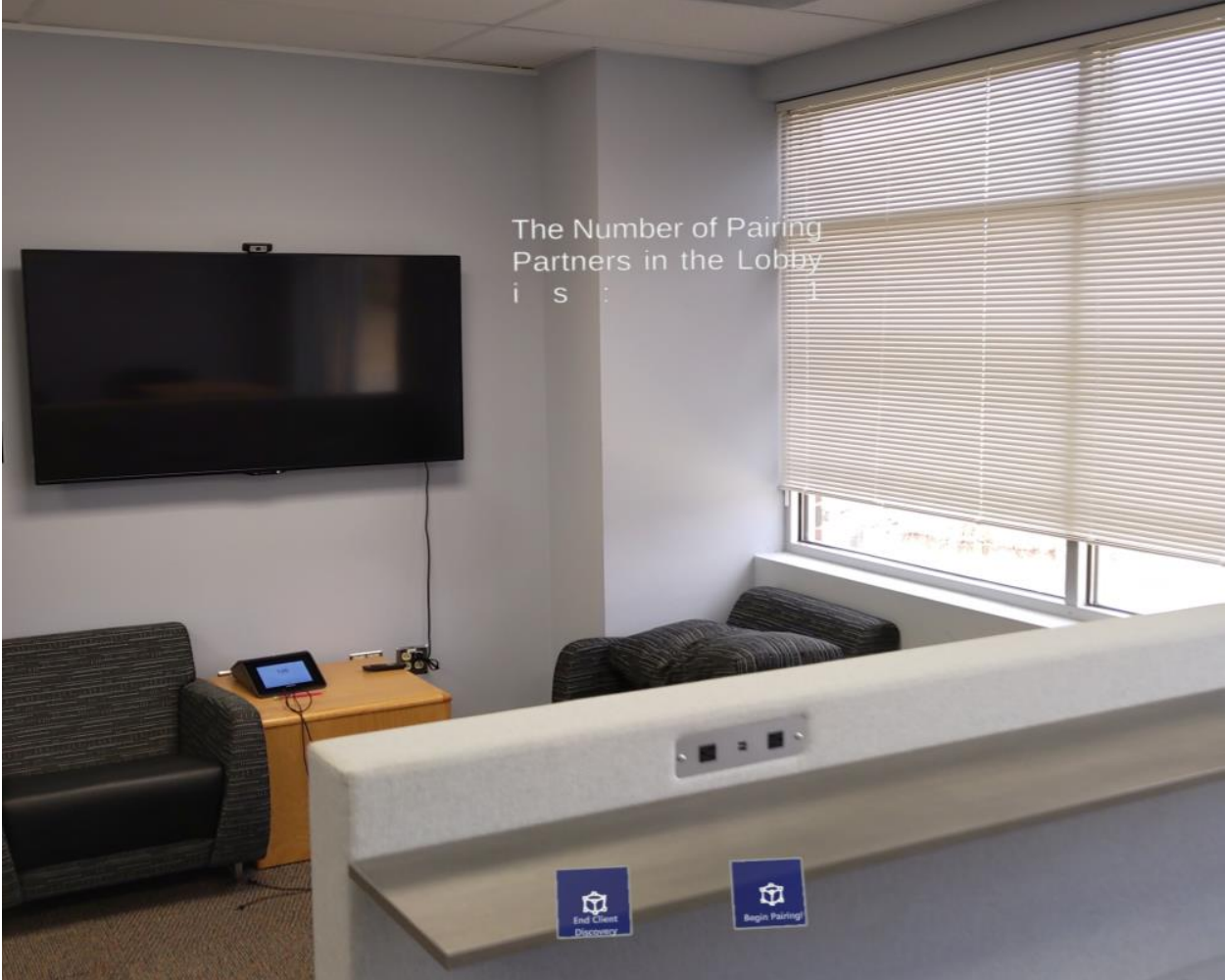}
         \caption{Host/Client identification}
         \label{fig:GazePairExamplea}
     \end{subfigure}
     \hfill
     \begin{subfigure}{0.26\textwidth}
         \centering
         \includegraphics[width=\textwidth]{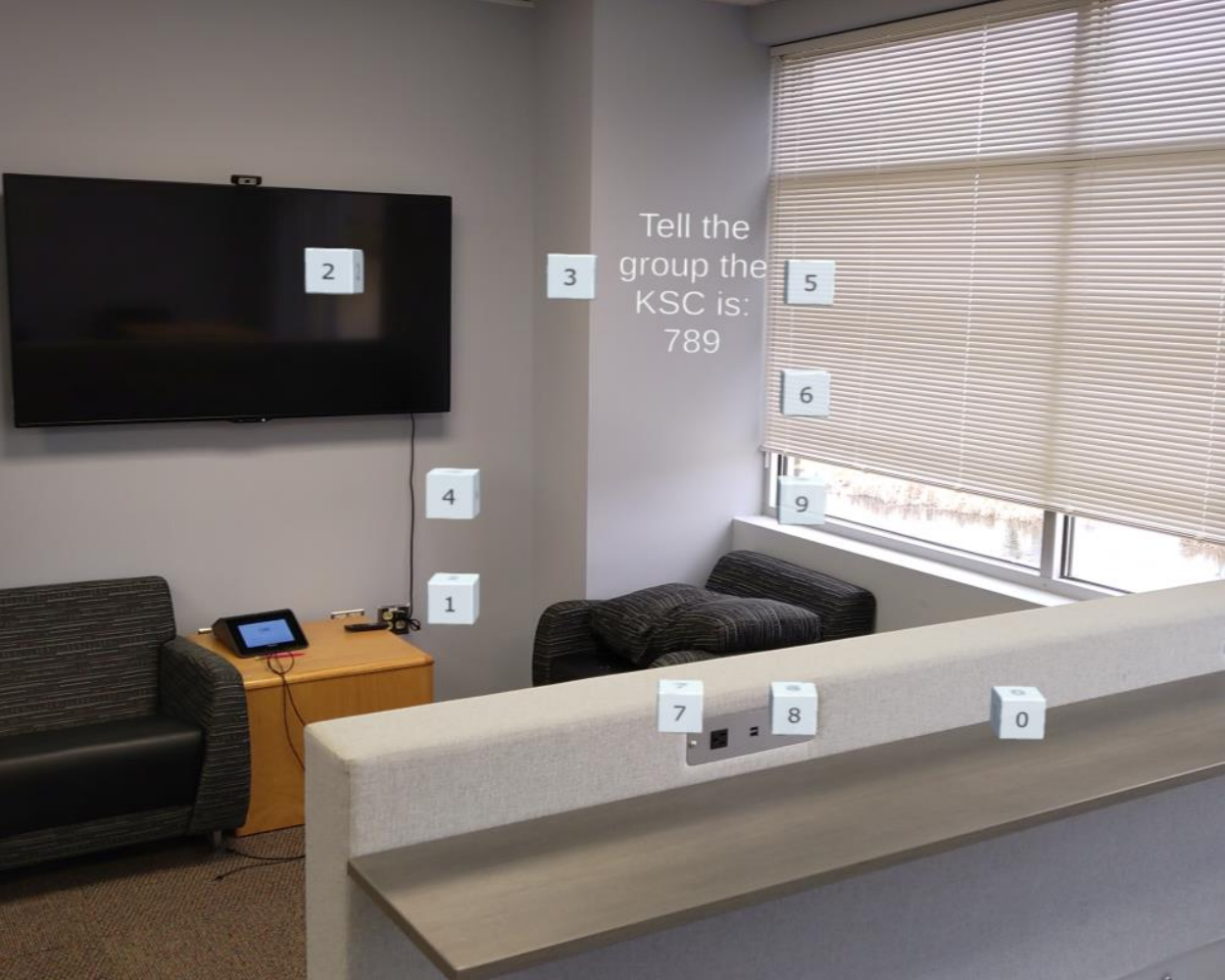}
         \caption{Host's view of the pairing}
         \label{fig:GazePairExampleb}
     \end{subfigure}
     \hfill
     \begin{subfigure}{0.26\textwidth}
         \centering
         \includegraphics[width=\textwidth]{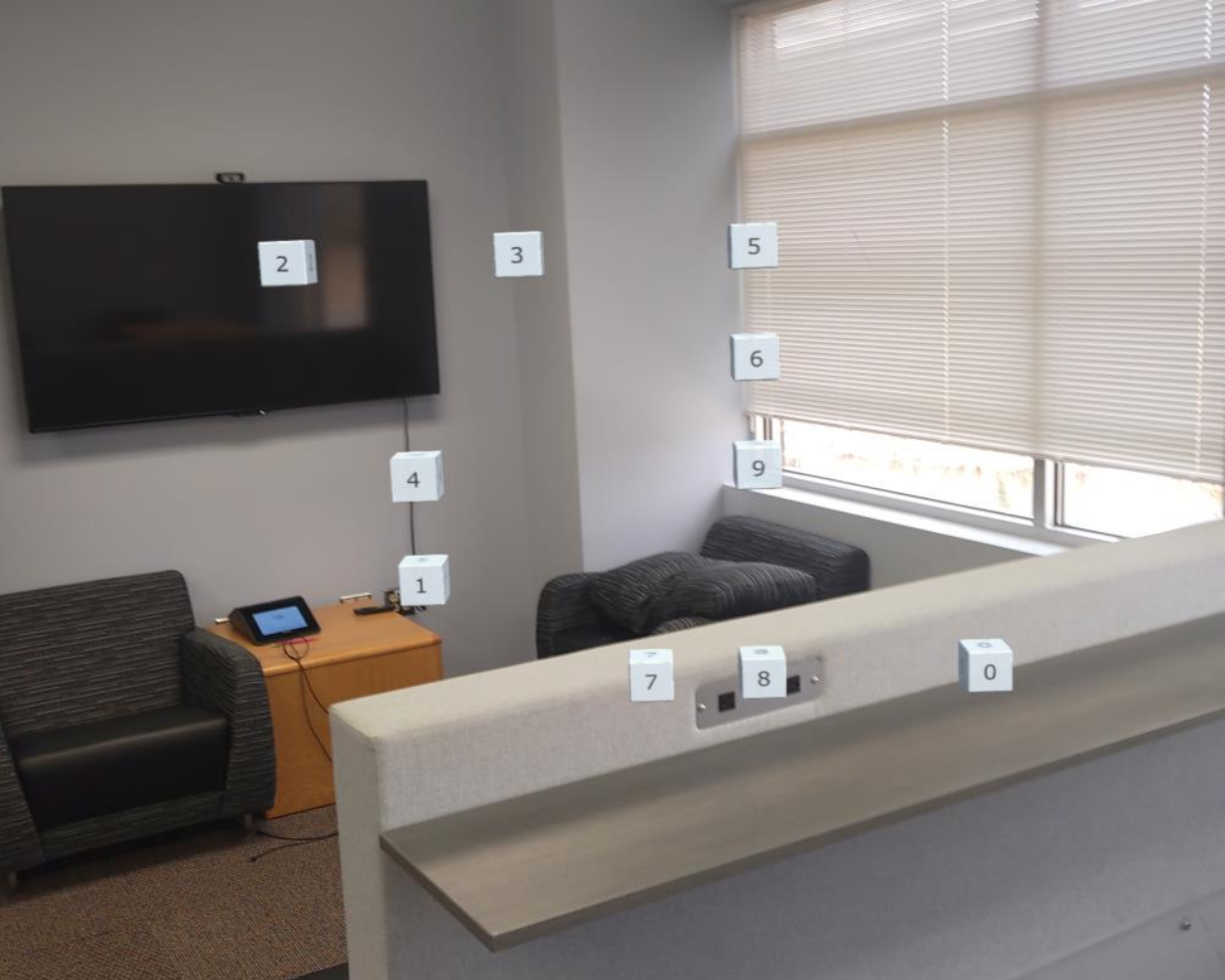}
         \caption{Client's view of the pairing}
        \label{fig:GazePairExamplec}
     \end{subfigure}
        \caption{An illustration of Steps 1 and 2 of GazePair. (a) shows the first scene where the host and client select their applicable roles and the pairing protocol establishes their relationships. (b) and (c) show the view of the pairing scene from both the host's and client's points of view, respectively.}
        \label{fig:GazePairExample}
\end{figure*}

\begin{figure}
\centering
\includegraphics[scale=.5]{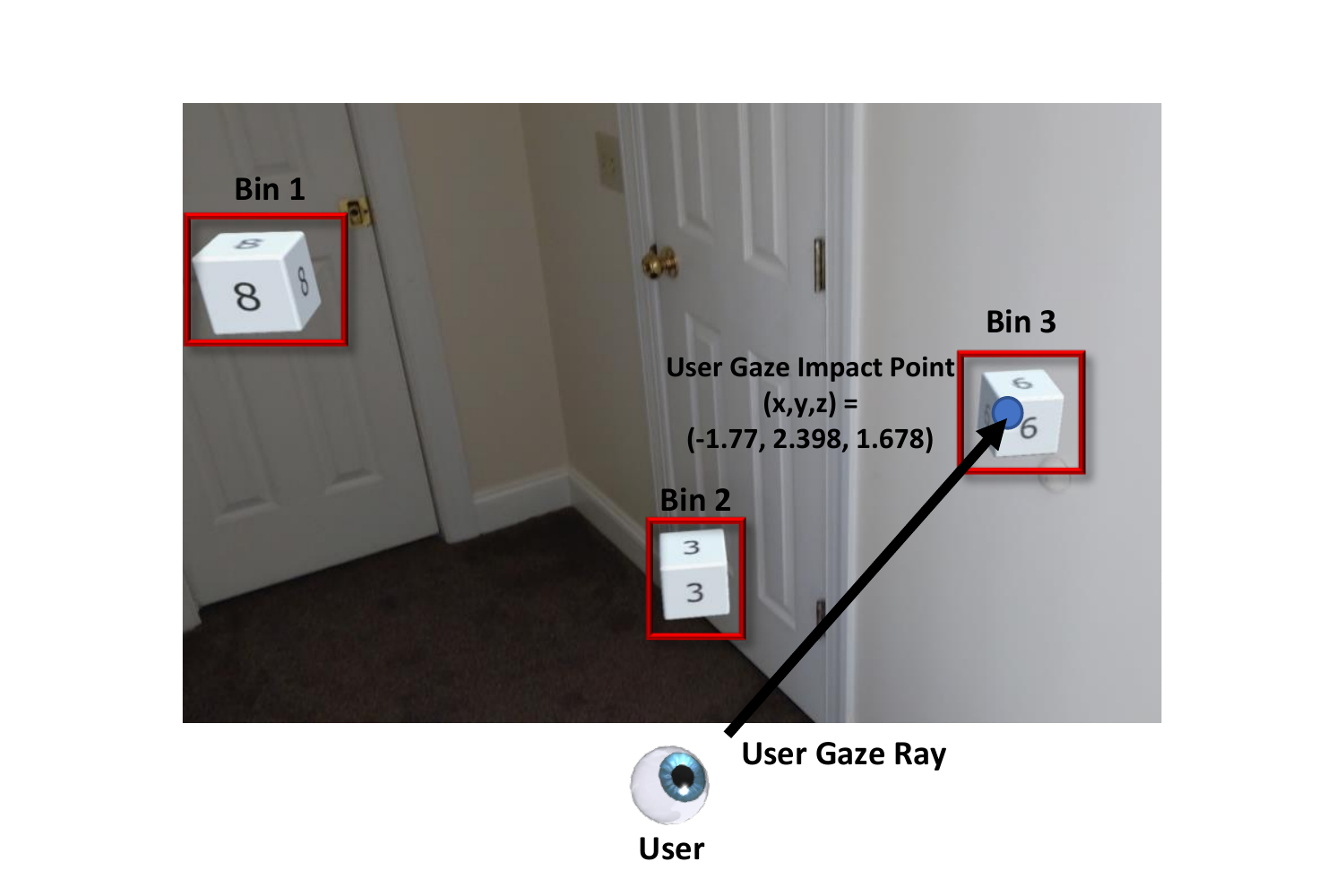}
\caption{An example of the eye gaze discretization technique used in GazePair. After a KSC is spoken to each client from the host, each user finds the required numerically-labeled hologram, gazes at it, signals the protocol to take a snapshot of the gaze location, and moves on to the next required hologram. Each gaze sampling is discretized and becomes a part of the shared secret. In this example, the first gaze sampling is taken when the user gazes at the hologram labeled ``6''. The raw gaze value is (-1.770, 2.398, 1.678), which is then discretized to ``-222'', based on the discretization parameters, and added to the shared secret. The length of this value will increase as the user takes more gaze samplings directed by the KSC.}

\label{fig:BinningExample}
\end{figure}

\subsubsection{Host/Client Identification}
GazePair creates a host/client relationship between two or more HMDs. For testing, we hard-code the IP address of the host on all HoloLens~2 HMDs, and make a single HoloLens~2 HMD the dedicated device for initiating the pairing procedure, to ensure mistyped IP addresses do not alter the testing data. In future implementations, either method can be used. Fig.~\ref{fig:GazePairExamplea} shows the view of the host and the client in this stage.

The host, and the host only, is able to monitor the number of clients registered by MLAPI as present in the pairing lobby. This serves to ensure that the host has made a connection with the number of clients they expect, and also serves to ensure that, if clients additional to the number expected join the pairing instance, the host is informed. This will not stop a surreptitious MITM-style attack but will mitigate the risk of a client accidentally joining the incorrect pairing instance by allowing the host visibility of how many clients are connected to their pairing session.

\subsection{Gaze and Entropy Extraction}\label{sec:GazeEntropyExtraction}
After the creation of the host/client relationship between two or more HoloLens~2 devices, the host randomly distributes 10 numerically-labeled holograms (denoting digits from 0 to 9) within a given visibility threshold. These numerically-labeled holograms are placed no closer to each other than twice the distance of a given error threshold in any direction. Attention is paid to ensuring that these numerically-labeled holograms are created in user-friendly locations, and not on top of users, behind users, or too far from users. An example of this from the host's and the client's viewpoint is shown in Figs.~\ref{fig:GazePairExampleb} and \ref{fig:GazePairExamplec}, respectively. The host then speaks a randomly-generated KSC of a given length.  Each user, independently, selects the numbers specified by the KSC, one at a time using eye gaze. The selection is done using MRTK's GazeProvider APIs~\cite{MRTK, MRTKEyeTracking}, to access a gaze ray, from the user's eyes to the hologram, striking at a given point.  Once the user's gaze is placed on the intended hologram, the user takes a snapshot of the current gaze location by using the ``air tap'' gesture~\cite{AirTap}.  

\begin{remark}
\label{rem:airtap} 
The ``air tap'' gesture is used jointly with the user's gaze location to harness gaze data. This gesture requires a user to ``pinch'' in the air to signal the device to select an eye gaze target for action. This gesture does not require the user to direct this ``tap'' at any given hologram, only that the gesture is visible to the HoloLens~2's Articulated Hand Tracking (AHAT) short-throw camera~\cite{Ungureanu2020HoloLens2R}, which has a field of vision several feet outside what the user can see through the HoloLens~2's visor. 
\end{remark}

As part of the prototype testing, we also implemented a gesture-less method of gaze collection. This method asks a user to simply keep their gaze on an intended hologram, and after an elapsed period (e.g., two seconds), GazePair records the gaze position and discretizes this portion of the shared secret. This method is also effective, but vulnerable to novice AR users staring too long at a given hologram while they are learning the GazePair system. We have chosen the ``air tap'' as it is a way to ensure an extremely low error rate in gaze collection, improving the pairing success rate.

The user's gaze location is then discretized shown in Fig. ~\ref{fig:BinningExample} to allow for effective shared secret generation and error correction. Given an error threshold, GazePair creates a grid of 3D spaces covering all possible hologram locations. Since the host generates the shared holograms at specific locations at the center of these ``error cubes'', and given the error threshold, it is not possible for a user to select a given hologram and produce the discretized location of another. This allows GazePair to ensure discretization within a given threshold without having to publicly exchange discretization parameters as in Tap-Pair~\cite{10.1145/3374664.3375740}. Each discretized value becomes a part of the shared secret, and each value is concatenated to become the final string used as the input to the Password-Based Key Derivation Function v2 (PBKDF2) for symmetric key generation. If the users enter an incorrect key, the shared secrets and symmetric keys will not match, and the pairing will fail. This shared secret, once successfully generated, is kept local to each device and will never be shared across the local network.

\begin{remark}
During the pairing process, users are allowed and encouraged to move freely around the room.  There is no requirement in GazePair to sit, stand, or remain stationary. In fact, some users find it more entertaining to alter the way they view the numerically-labeled holograms, while some prefer to sit.  Either is acceptable and did not alter GazePair's efficiency or effectiveness during evaluation. 
\end{remark}

\subsection{Key Generation}\label{sec:KeyGeneration}
After the establishment of the shared secret, each instance of GazePair creates a 256-bit AES key using the shared secret as an entropy source. A well-known and well-established technique to generate symmetric keys from passwords or other shared secrets is Password-based Key Derivation Function v2.1 (PBKDF2), as identified in IETF RFC 8018~\cite{rfc8018}. While known vulnerabilities exist in this method of key generation, including vulnerability to rainbow table attacks using advanced Graphics Processing Units (GPUs), GazePair requires potential pairing partners to commit to a single symmetric key during the pairing process, using unique and random initialization vectors and salt values, mitigating the problem of offline rainbow table attacks. 

\subsubsection{PBKDF2}
Accepted by the IETF in 2000 and updated in 2017, PBKDF2 seeks to protect a relatively low-entropy secret used to create keys, namely a password or other shared secret~\cite{rfc8018}. To do this, PBKDF2 hashes a password input with an initialization vector and adds artificial computational work to make the specter of a rainbow table or dictionary attack more difficult. This computational work, similar to Bitcoin's Proof of Work concept for validating blockchains~\cite{BitcoinProofOfWork}, is intended to increase the computational cost of hashing all possible passwords a user might input. If the attacker knows the initialization vector, they must then begin the computationally expensive process of generating all possible values of the password and wait the time required for the artificial computational work defined by the number of iterations. As recently as 2021, PBKDF2 has been proposed for sensitive systems, such as cyber-physical systems, with modifications to increase the computational resources required to generate password digests~\cite{10.1145/3408310}. For GazePair, we choose to use the SHA256 hashing function for PBKDF2 due to its collision resistance and high security, and a total of 50,000 iterations. We choose 50,000 iterations as it is a number that creates no noticeable performance degradation on the HoloLens~2, while larger values create noticeable and unacceptable degradation.

Both a salt and initialization vector are required for PBKDF2 key generation and AES encryption and decryption. Using the Microsoft .NET pseudo-random number generator~\cite{NetPRNG}, the host generates a 64-bit random number used as the salt and initialization vector. These values are published to all clients, ensuring that the salt and initialization vector are both random and known to all.

\subsubsection{Analysis of Generated Entropy}\label{sec:entropy_analysis}

Using the accepted method of calculating password entropy, denoted by $E$, from the total number of shared secret possibilities, denoted by $S$, specifically we calculate entropy as follows:
\begin{equation}
    E = \log_2{S}. \label{eq:entropy}
\end{equation}

Assume a total number of $K$ numerically-labeled holograms. Each hologram has an $(x,y,z)$ value on the 3D Cartesian plane, and these values are limited to the value pool, i.e., the total possible values of each axis, denoted by $X_t$, $Y_t$, and $Z_t$. Each $(x,y,z)$ value is non-repeating for each combination to prevent two numerically-labeled holograms from spawning in the same location. We use the permutation below to solve for the total number of possible key arrangements of the total $K$ numerically-labeled holograms, denoted by $N_K$: 
\begin{equation}
    N_K = {X_t \cdot Y_t \cdot Z_t \choose K} \cdot K! = \frac{\left(X_t \cdot Y_t \cdot Z_t\right)!}{\left(X_t \cdot Y_t \cdot Z_t - K\right)!}.
\end{equation}
The numerically-labeled holograms selected are dictated by the order and value of the KSC, which is generated by GazePair and spoken by the host. The number of numerically-labeled holograms selected is the length of the KSC, denoted by $P$. Since the KSC has non-repeating digits (to aid in input and error detection), the total number of KSCs, denoted by $N_P$, can be calculated as follows:
\begin{equation}
    N_P = {K \choose P} \cdot P! = \frac{K!}{\left(K - P \right)!}.
\end{equation}
Hence, the total number of shared secret possibilities is $S=N_K \cdot N_P$.
Plugging it into Eq.~\eqref{eq:entropy}, we calculate entropy $E$ as follows:
\begin{equation}
    E = \log_2{S} = \log_2{(N_K \cdot N_P)}.
\end{equation}

Specific to GazePair, we calculate entropy slightly differently, as all the numerically-labeled holograms are on the same Z-axis to ensure visibility, and thus, there are fewer permutations. We also use a KSC length of 3 digits, further explained in Section~\ref{sec:Discussion}. Additionally, we do not spawn numerically-labeled holograms at (0,0) on the $(x,y)$ planes, so as to not obscure the host's view of the KSC prompt. In GazePair, we use $X_t=7$, $Y_t=6$, and $Z_t=5$. We also have a total of 10 numerically-labeled holograms, 0-9, so $K=10$. We have $P=3$, as the KSC has 3 digits, which are non-repeating. Hence, we can calculate the following:

\begin{equation}
\begin{split}
   N_K &= {X_t \cdot Y_t - 1\choose K} \cdot Z_t \cdot K! = \frac{(X_t \cdot Y_t - 1)!}{(X_t \cdot Y_t - 1 - K)!} \cdot \\Z_t &= \frac{\left(7 \cdot 6 - 1\right)!}{\left(7 \cdot 6-1 - 10 \right)!} \cdot 5 = 2.034 \times 10^{16}, 
   \\N_P &= \frac{K!}{(K-P)!} = \frac{10!}{(10-3)!} = 720~(\text{9 bits of entropy}),
   \\E &= \log_2 (N_K \cdot N_P) \bold{\approx 64~\text{bits of entropy}}.
\end{split}
\end{equation}

\section{Experimental Evaluation}
\label{sec:evaluation}
One-to-one pairing tests take an average of 9.02 seconds, with a 98.3\% pairing success rate. One-to-two pairing tests take an average of 12.58 seconds, with a 96.6\% success rate. This is accomplished using a prototype that is deployable on the breadth of AR devices that use eye gaze tracking and MRTK, and with a design that is deployable on the breadth of MR devices using  any form of gaze tracking.

\subsection{Software and Hardware Setup}
We built the GazePair prototype in Unity 2020.3.16f1, using MRTK version 2.7.2 and MLAPI version 0.1.0. GazePair was deployed on Microsoft HoloLens 2 HMDs running Windows Holographic for Business Build 20348.1438. Blender version 3.0.1 was used to create the custom numerically-labeled holograms, and all source code was completed in C\#. Our prototype is posted as open source~\cite{GazePairGit}.

\subsection{Evaluation Design}
The GazePair prototype was tested by 20 participants with varying ages, vision capabilities, and technical backgrounds. The entirety of the evaluation was approved by the University's IRB. Participants were recruited using the University's graduate student email distribution list, or from the local, non-student population. Upon arrival at the test site, each participant signed a consent form and completed a basic demographic questionnaire. Each user was then given a brief (ten minutes) tutorial on the basic operation of the HoloLens~2, including adjustment of the fit of the HoloLens~2 HMD, accessing the main menu, AR gestures, and operating GazePair. Additionally, we asked each user to complete an eye calibration, as suggested by Microsoft, for each user before every set of tests. The participants were allowed to remain stationary during the tests, or move freely, as they felt comfortable. For each group of participants, a host was randomly selected and given the HoloLens~2 with the hard-coded IP for the host, while the other participants served as the client. All tests were conducted over a Netgear AC1750 router without an internet gateway, to ensure that all tests were over a local network. Additionally, at the end of the testing session, the participants completed a usability survey, with responses on a 5-point Likert scale ranging from ``Strongly Disagree'' to ``Strongly Agree''~\cite{Likert}.

Below are the metrics measured for each test iteration.
\begin{list}{\labelitemi}{\leftmargin=1.5em \itemindent=-0.0em \itemsep=.2em}
    
    \item \textbf{Total time required to complete pairing.} The GazePair logging script \cite{CSVLogger} begins a timer at the moment that the host is satisfied with the number of participants in the pairing lobby and begins the pairing protocol.  From that moment, the elapsed time is measured until the completion of the KSC entry of the final participant to finish the sequence. 
    
    \item \textbf{Success rate.} GazePair records the boolean value that corresponds to the host's ability to decrypt messages from all clients in the pairing protocol. If any message cannot be decrypted by the host, the pairing protocol is considered a failure, and the host is notified.
    
    \item \textbf{Number of pairing partners.} The logging script records, for each test, how many participants were present for the pairing session.
\end{list}

\subsubsection{Demographics}
Our testing consists of 6 self-described women and 14 self-described men, only four of which self-profess any experience with AR devices. The ages of the participants range from 20 to 72. Users have a wide variety of vision capabilities, ranging from perfect vision without glasses or contacts to impaired with vision correction. Users come from backgrounds that vary from computer science graduate students, military service members, to retirees. 

\subsubsection{Evaluation Metrics}

Additionally, we chose to evaluate the ability of GazePair to satisfy its design objectives with a series of tests and evaluations. Specifically, we evaluate the GazePair design using the following experiments.

\begin{list}{\labelitemi}{\leftmargin=1.5em \itemindent=-0.0em \itemsep=.2em}

    \item \textbf{Success Rate.} We evaluate GazePair's effectiveness by analyzing the number of successful tests against the total tests, and the reasons for each failed test. This is done for both one-to-one and one-to-two pairing attempts.
    
    \item \textbf{Usability.} We evaluate the time required to complete a pairing iteration in order to establish evidence of its efficiency. We discriminate between one-to-one and one-to-two pairing attempts. 
    
    \item \textbf{The usability over time.} As a hypothesis, we expect usability to increase as users increase their familiarization with the system. We present pairing times, compared against the number of pairing attempts completed, as a way to analyze this.
    
    \item \textbf{GazePair vs. existing solutions.} We evaluate GazePair qualitatively and quantitatively against all known AR pairing solutions.
\end{list}

\subsection{Pairing Performance}

\emph{With 240 tests, 97.5\% of pairings complete successfully.} The six tests that failed were due to misselected numerically-labeled holograms, either selected in the incorrect order or due to misunderstanding the KSC. \emph{For one-to-one pairing attempts, 118 of 120 (98.3\%) are successful, while 96.6\% of the 120 one-to-two pairing attempts succeed.} 
\emph{The one-to-one pairing performance is an 8\% improvement on the most current and advanced AR device pairing technique, while the 96.6\% pairing success rate of the one-to-two tests shows GazePair's scalability.} 

\begin{figure}[!t]
\centering
\includegraphics[scale=.35]{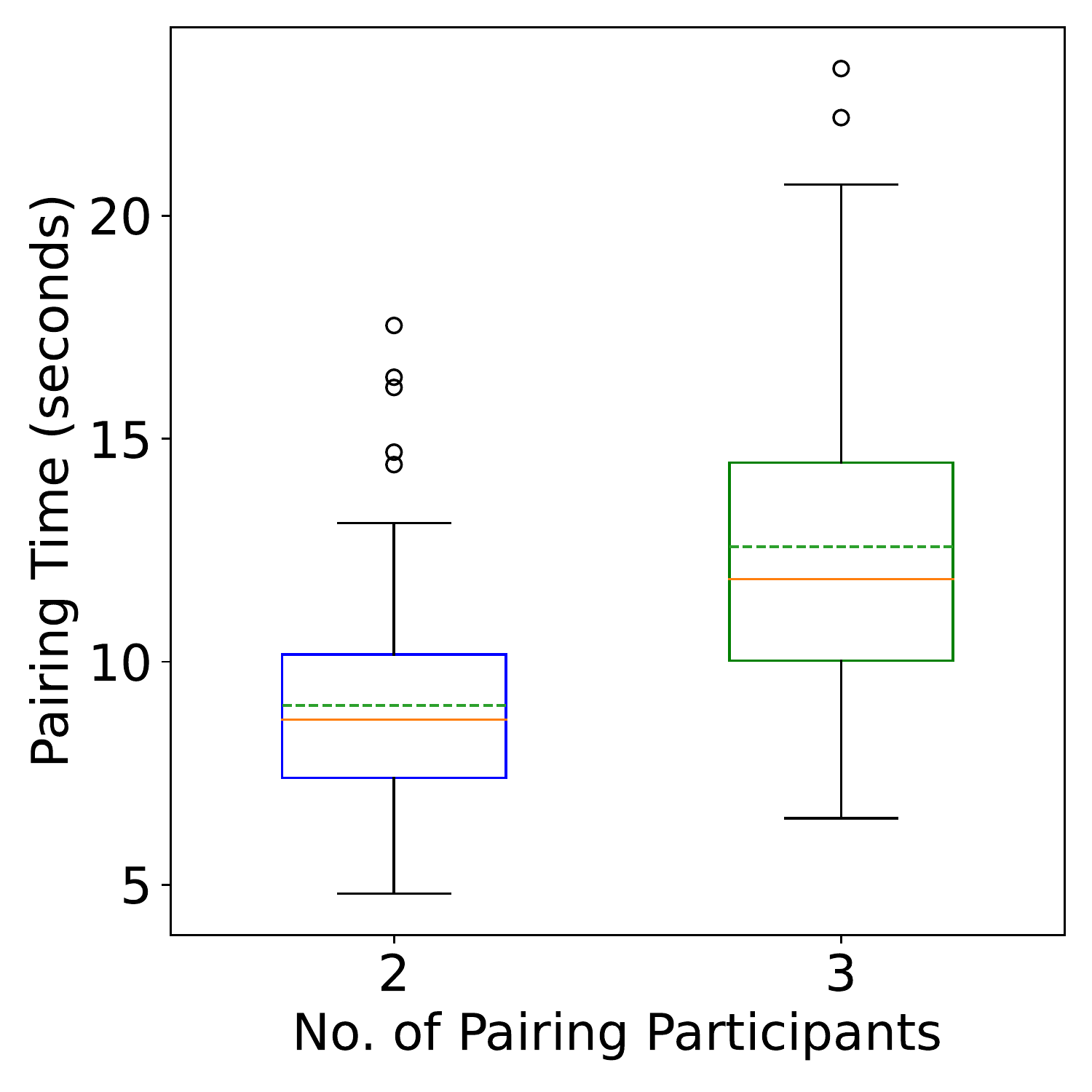}
\caption{A boxplot of the total pairing time, by the number of pairing partners. The median time is shown with the solid, horizontal, orange line.  For one-to-one pairing, the median time is \high{8.7} seconds; for one-to-two pairing, the median time is \high{11.83} seconds. The average time for each is represented by the dashed, horizontal, green line. The average time for one-to-one pairings is 9.02 seconds; for one-to-two pairings, the average time is 12.58 seconds.}
\label{fig:BoxPlot}
\end{figure}

\begin{figure}[!t]
\centering
\includegraphics[scale=.40]{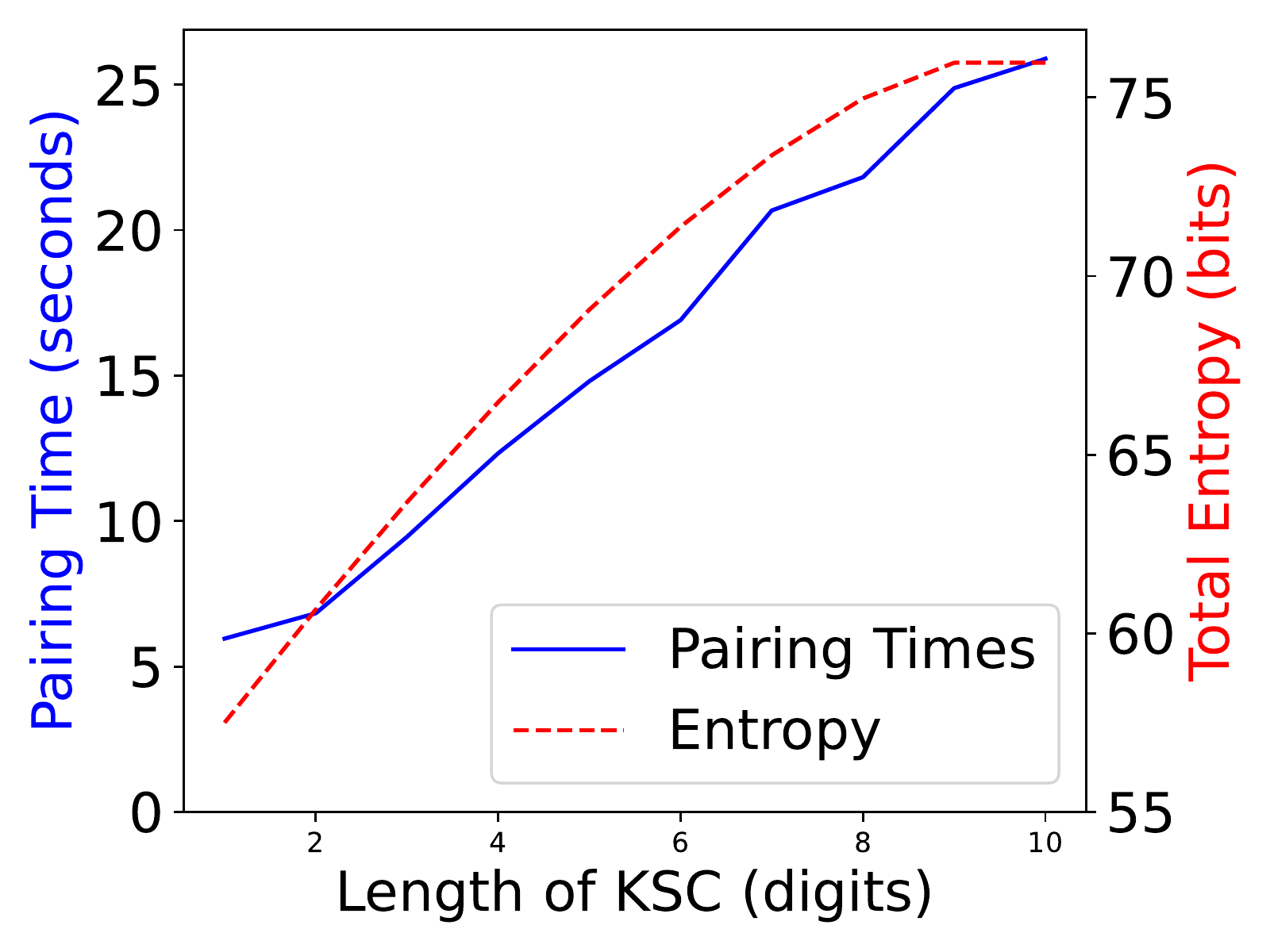}
\caption{An illustration of the tradeoff between the length of the KSC, the time required to complete one-to-one pairings, and the generated entropy. Note the generally linear increase in pairing times relative to the KSC length.}
\label{fig:KSCvsTime}
\end{figure}

Additionally, we compare the times to complete a one-to-one pairing with the length of the KSC. Fig.~\ref{fig:KSCvsTime} shows both the average time to complete a pairing attempt at each possible KSC and also the resulting entropy. This shows that the pairing time generally increases as the length of the KSC increases. We also note that as the KSC began to increase, participants were no longer able to speak the KSC in a single phrase and began to divide the KSC up into parts.

\subsection{Usability}\label{sec:usability}
Fig.~\ref{fig:BoxPlot} presents a box plot of all the pairing tests conducted, by the number of pairing partners. \emph{The average time to complete one-to-one pairings is 9.02 seconds}, from the host initiating the pairing protocol to the host's successful decryption of each client's ciphertext using the host's AES key. For one-to-two pairings, the average pairing time increases to 12.58 seconds. We believe that this shows the efficiency of the GazePair system, as measured by user time requirements. GazePair matches or improves on the pairing time of the three known local AR pairing solutions. 
As noted in Remark~\ref{rem:airtap}, we also implemented a gesture-less version of GazePair using gaze dwell as the trigger to collect the gaze data. This technique is potentially even more user-friendly, but the likelihood of novice AR users accidentally selecting the incorrect holograms while learning AR gestures increases. While this would be lessened with experience, we choose to use the ``air tap'' due to the much lower occurrence of mis-selected numerically-labeled holograms by novice AR users.

\begin{figure}[!t]

     \centering
     \begin{subfigure}{0.35\textwidth}
         \centering
         \includegraphics[width=\textwidth]{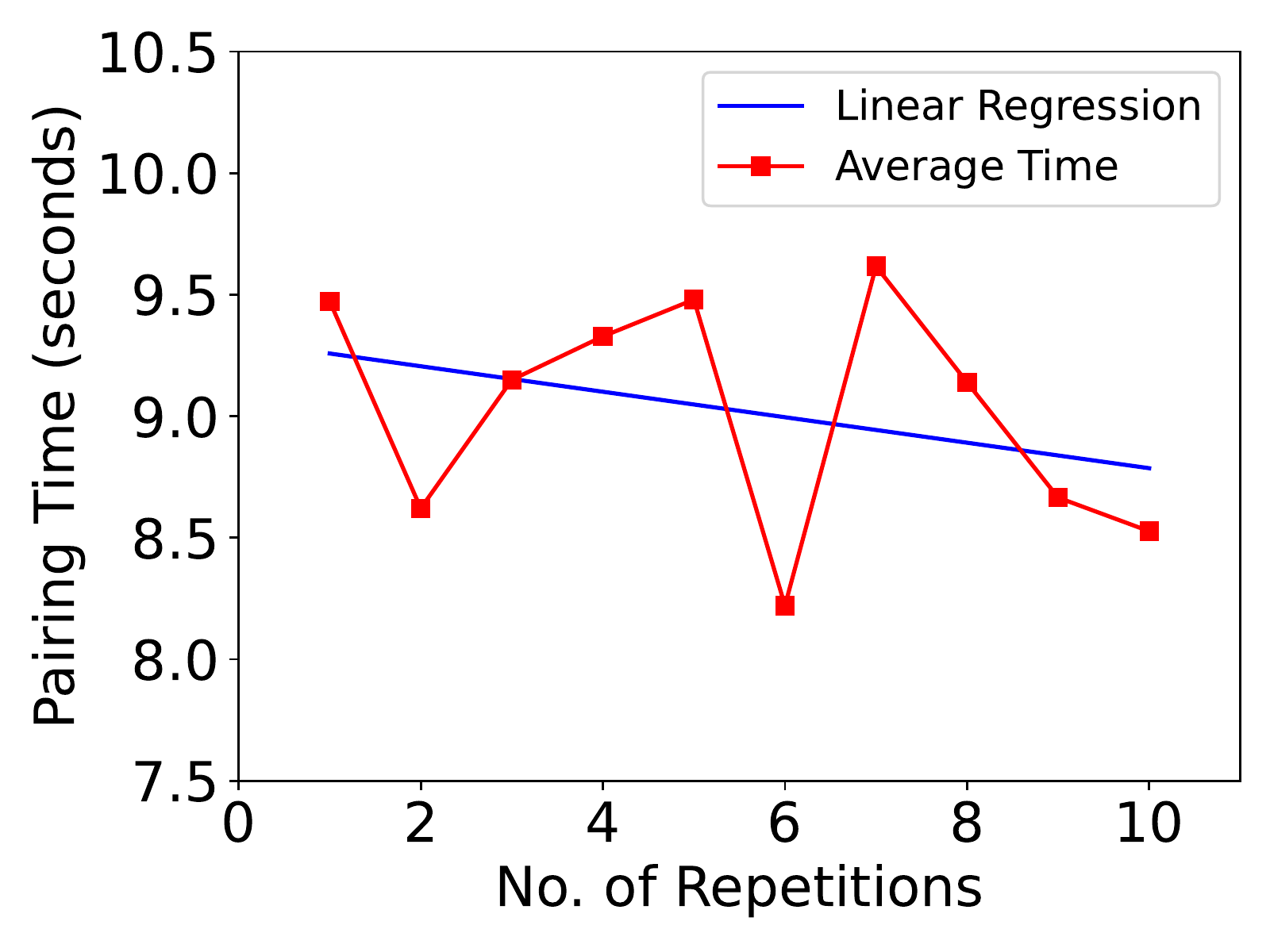}
         \caption{One-to-one pairing}
         \label{fig:LearningEffect2Participants}
     \end{subfigure}
     \begin{subfigure}{0.35\textwidth}
         \centering
         \includegraphics[width=\textwidth]{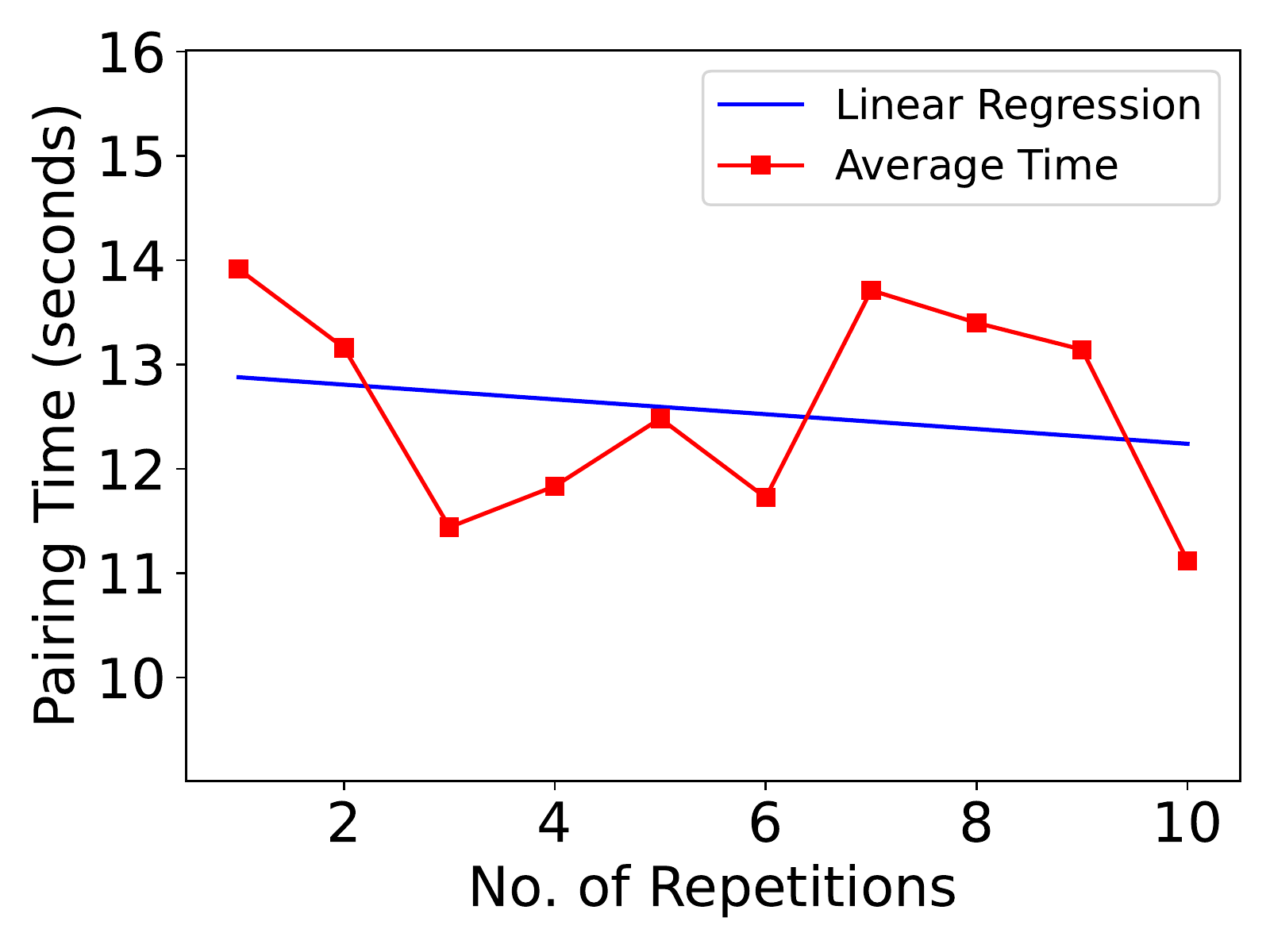}
         \caption{One-to-two pairing}
         \label{fig:LearningEffect3Participants}
     \end{subfigure}
     \hfill
        \caption{The effect of multiple repetitions on the pairing time shown by average time and a linear regression of the time required for sequential tests. (a) shows the results of participants in one-to-one pairing. (b) shows the results of participants in one-to-two pairings. The generally downward trend in the pairing time indicates that as users begin to become familiar with the system and AR gestures, the system becomes more usable.}
        \label{fig:LearningEffect}
\end{figure}

Additionally, we evaluate the usability of the system based on the feedback from participants at the end of the testing based on the 5-point Likert scale. Of the 20 participants, 95\% stated that they ``Strongly Agreed'' or ``Agreed'' the system was easy to use, and that they believed most users would learn the system ``quickly''; 90\% responded that they ``Strongly Disagreed'' or ``Disagreed'' the system was cumbersome. We believe that this user feedback further reinforces the usability of the GazePair design.

\subsection{The Usability Over Time}\label{sec:EfficiencyOverTime}

Fig.~\ref{fig:LearningEffect} presents the time required to complete a test, regardless of pairing success, over multiple pairing iterations. The generally downward trend of the pairing time, as the familiarity with the system increases, shows the effect of experience with the system.  As the participants spend more time with GazePair, their familiarity with AR increases, and they are able to more easily complete pairing attempts. Specifically, the average pairing time for the first attempt at a one-to-one pairing is 9.47 seconds, falling to 8.52 seconds by the tenth. Similarly, the average first attempt with three participants in a one-to-two pairing is 13.9 seconds, falling to 11.09 seconds by the last attempt.
Given the relative inexperience using AR of the participants, this appears to be expected. Additionally, we believe this speaks to the usefulness of eye gaze. Users, regardless of AR experience, intuitively understand how to gaze at an object. No testing participant expressed confusion over the use of eye gaze.

\subsection{GazePair vs. State-of-the-Art Solutions}
Table~\ref{tab:researchComparison} gives a comparison of GazePair to the state-of-the-art research into AR pairing. LGTM is useful as a baseline solution, but it is not usable on any known AR device given its requirement for wireless localization, a requirement that also contributed to its low success rates.  Additionally, its facial recognition will be frustrated by AR devices with visors obscuring the user's face. HoloPair's success rates and time requirements are on par with GazePair's, but HoloPair is hampered by its requirement for tracing and waving between two users and the ability of these users to simply ``skip'' the pairing attempt. Additionally, HoloPair's generated entropy is about 26 bits in its most complex variation, something the authors claim is ``not large''. For usability, GazePair and Tap-Pair are roughly equivalent. Both require user gestures. TapPair requires a user to direct their head toward a place designated by the pairing initiator, designated by a finger or post-it, and ``tap''. GazePair requires vocal communication as well as using the ``air tap'' gesture. Also, similar to Tap-Pair, the authentication system in~\cite{InvestigatingTheThirdDimension} leverages the same insight as our system - AR users have their unique knowledge of the virtual objects in the scene, but also produces low entropy and high failure rates in evaluation scenarios. Compared with Tap-Pair and similar authentication systems, \emph{GazePair increases the success rate of pairing attempts by 8\%, and quantifiably proves its scalability with more than two users, something only hypothesized in Tap-Pair.}  GazePair also improves on the 9-11 bits of total entropy generated by Tap-Pair, by adding randomness in the location of holograms and in the choice of the KSC. In addition, the work in~\cite{InvestigatingTheThirdDimension} uses physical controllers, which allows attackers to observe the input behaviors and guess the password (in 12.5\% to 18.5\% of attempts). Differently, we use eye gaze, which is invisible to attackers, so we can provide better defense against an attacker with the ability to observe the pairing. Finally, GazePair's design is applicable to the breadth of MR devices that use eye gaze tracking, and the prototype itself is deployable on any device harnessing eye gaze using Microsoft's MRTK, something no other solution claims.

\section{Discussion}
\label{sec:Discussion}

\subsection{Availability of Gaze Tracking Sensor}
The gaze tracking sensor is essential for supporting GazePair. Through a survey of documentation related to current or upcoming AR HMDs, we find that most of the popular MR headsets already (or will) support gaze tracking. For example, the AR headsets of Magic Leap, Magic Leap~1, support gaze tracking in their current model, and this feature will continue to be available in their upcoming new models~\cite{EyeGazeComingDeviceNearYou}.  In addition, Meta (formerly Facebook) earlier stated that they would include sensors for both face and eye tracking on their Oculus Quest Pro~\cite{Oculus}. This shows that the gaze tracking sensor has already been and will continue to be essential hardware on future MR headsets, indicating that GazePair can be easily implemented on future MR devices without introducing any extra overhead.

\subsection{Length of the KSC}
In the GazePair evaluation, we used a KSC length of 3 digits. This was a compromise between the time required to input the KSC and the additional entropy provided by longer KSC lengths. For example, at a KSC length of 10 digits, GazePair can provide 76 bits of entropy. We chose 3 digits, as this is the smallest KSC that provides entropy deemed to be ``strong''~\cite{MinimumEntropy} while remaining usable in terms of pairing time compared to existing solutions.

\subsection{Applicability to the Metaverse}
As discussed in Section \ref{sec:intro}, the advent of the Metaverse is a motivational factor behind the expansion of AR. Large corporations, such as Meta, envision this digital collaboration space as the future of work and entertainment~\cite{Meta}. Certainly, the Metaverse could affect our daily lives in the near future. However, these large environments are envisioned to be hosted and implemented on centralized servers, likely under the control of large corporations such as Meta. Data collected can include visual and audio recordings of the user, iris data, body type, and movement style. What if a small organization, such as a social group, does not wish to have data like this transferred across these large servers in far-away places? Such small groups could use ad-hoc, local, secure pairing techniques such as GazePair to transfer spatial data, visual and audio data, and the like while not having to worry about corporate data collection or misuse. 

\subsection{GazePair's Performance Under Threat}\label{sec:Attack}
To clarify GazePair's performance under attack, we discuss the ability of an attacker to compromise the security offered by GazePair in each of the proposed threat conditions. If the attacker has access to GazePair's network traffic, they can use this information to deduce the location of each numerically-labeled hologram and the IV/Salt values. This is the most advantageous attack vector as the only barrier to compromise of the symmetric encryption keys is the knowledge of the KSC. Given a KSC length of 3, the attacker has a 0.1\% chance to correctly guess the KSC, or a total of 9 bits of overall entropy as outlined in Section~\ref{sec:entropy_analysis}. This chance is as low as 0.00003\% with a KSC length of 10.

In the second threat condition, where the attacker is co-located with the pairing participants but does not have access to GazePair's network traffic, the challenge for the attacker increases. As outlined in Section~\ref{sec:entropy_analysis}, GazePair randomly generates all 10 numerically-labeled holograms in three dimensions. First, GazePair chooses a randomly selected $z$ value (i.e., depth) from 5 possible choices. Then, all 10 numerically-labeled holograms are randomly generated on a $7 \times 6$ grid at this depth. Assuming a KSC length of 3, the attacker would need to guess from 344,400 possible combinations of holograms. Even assuming this correct guess, the attacker does not know the IV/Salt values shared between devices, making the knowledge of the shared secret irrelevant.

As mentioned in Section~\ref{sec:threat}, we assume that the attacker cannot simultaneously have knowledge of the KSC, the location of the numerically-labeled holograms, and the IV/Salt values. If this were not the case, GazePair's security could potentially be compromised.

\subsection{Limitations of GazePair}\label{sec:limitations}
We have shown that GazePair is an efficient, effective, and safe AR device pairing technique. Even so, there are certainly areas for improvement. We summarize GazePair's limitations as follows.

 \begin{list}{\labelitemi}{\leftmargin=1.5em \itemindent=-0.0em \itemsep=.2em}
  
  \item \textbf{KSC entropy.} With the 3-digit KSC used during testing, the entropy of the KSC itself is only 9 bits. Even with PBKDF2 creating artificial work and increased computational time requirements to hash all KSC combinations, an attacker could potentially use the knowledge of the salt and initialization vector to test all possible key combinations against the known hologram locations if users do not use the generated symmetric encryption keys to transmit another stronger key. As shown in Section~\ref{sec:usability}, with an increased KSC length, while the entropy level improves, the pairing time also increases.
  
  \item \textbf{Out-of-band communication.} GazePair requires a spoken KSC between the host and the clients. As mentioned in Section~6.4, even hearing the KSC without knowledge of the Salt/IV or placement of the numerically-labeled holographic cubes does not give the attacker an advantage in this design. However, not all users may understand these technical details, and the act of speaking the KSC could indeed generate a perception of an insecure system, especially when the KSC is exchanged in public places. We believe that the potential misunderstanding about the safety of the system in the presence of other people/listeners could be alleviated with additional information about the high-level technical details. For example, if the users understand that the KSC cannot be used to compromise their pairing session unless the person overhearing their spoken communication can also intercept the Salt/IV and location of the individual holograms in 3D space, they may be much less concerned about this technique.
   
   \item \textbf{Expansion of paired group.} GazePair has no way to add participants to an established sharing group without completely re-attempting the GazePair protocol.  Keys are symmetric and generated uniquely per pairing attempt. If a user were to wish to participate in a given pairing group and was not present for the initial protocol, a new protocol instance would need to be initiated with current and new pairing partners.
\end{list}

\subsection{Future Work}\label{sec:futureResearch}
As noted above, GazePair's design still has areas for improvement. One potential area to improve on the entropy and out-of-band communication limitations is with spatial anchors~\cite{ASA}. Spatial anchors can allow multiple devices to not only see shared holograms but to see them in nearly the exact same location in the physical world, something not required by GazePair. Even so, transferring spatial anchors involves significant overhead as mentioned in Section~\ref{sec:intro}. Spatial anchors can be very large, and transferring them locally can be difficult. However, these anchors can allow users to interact with shared holograms at absolute locations, creating an opportunity to use this for new and more intuitive methods to pair devices without using a KSC. Specifically, these anchors can potentially allow users to physically locate the intended pairing partners in a room, but without the wireless localization requirement of HoloPair. This enables new ways to authenticate users by verifying their physical location. However, optimizing the transfer of spatial anchors across a local network should be a subject of future research to make this technique more viable and efficient.

\section{Conclusion}
\label{sec:conclusion}
In this paper, we designed a novel AR device pairing system, GazePair, which
leverages eye gaze tracking, a new and powerful AR technology. GazePair achieves efficient, secure, and scalable local AR device pairing with minimal user interaction while protecting user gaze data. This requirement has become increasingly self-evident as the prevalence of AR devices increases, and the need to share holographic information locally becomes more pressing. Using Microsoft HoloLens~2 HMDs, we implemented a prototype system of GazePair that allows two or more users, without an internet connection, to locally pair AR devices quickly and intuitively. Through experimental evaluation, we showed that pairing two or more AR devices using entropy created from gaze tracking can be efficient. Remarkably, users with minimal AR experience were able to efficiently and effectively use GazePair to pair two or more AR devices, achieving both a high success rate and a low pairing time that are an improvement on all existing solutions for AR device pairing. Furthermore, all of these can also be done without using techniques that limit the deployability of GazePair to a small subset of MR devices or jeopardizing user biometric data. 

With the advent of the Metaverse and future potential applications, such as in the military, healthcare, education, entertainment, and automotive manufacturing, the ubiquity of MR devices is only likely to increase. Techniques like GazePair are the building blocks to enable easy and efficient adoption of these MR devices by the broadest range of potential users in the broadest range of emerging MR applications.

\bibliographystyle{IEEEtran}
\bibliography{GazePair.bib}

\end{document}